\documentclass[11pt,a4paper,twoside]{report}
\usepackage{etoolbox}
\usepackage[dvips=false, pdftex=false]{geometry}
\usepackage[a4,center,font=sffamily]{crop}
\usepackage{footnote}
\usepackage[numbers,sectionbib]{natbib}
\def\ii\'{\i}
\def\ii{\'{\i}}

\def\na{-\kern-.4em\raise.8ex\hbox{{\tt \scriptsize a}}\ }
\def\naa{-\kern-.4em\raise.8ex\hbox{{\tt \large a}}\ }
\def\no{-\kern-.4em\raise.8ex\hbox{{\tt \scriptsize o}}\ }

\def\til{\scriptsize $\sim$ \kern-.4em}
\newcommand{\reell}{\kern+.23em\sf{1}\kern-.61em\sf{1}\kern+.76em\kern-.25em}

\usepackage{caption}
\DeclareCaptionFont{xipt}{\fontsize{11}{13}\mdseries}
\usepackage[font=xipt,labelfont=bf,indention=1.5em]{caption}
\usepackage{youngtab}
\usepackage{amsmath}
\usepackage{amssymb}
\usepackage{epsfig}
\usepackage{graphicx,color}
\usepackage{makeidx}
\usepackage{url}
\usepackage{tabu}
\usepackage[adobe-utopia]{mathdesign}
\usepackage[T1]{fontenc}

\usepackage{lineno}
\usepackage{indentfirst}
\usepackage{xcolor}

\usepackage[brazilian]{babel}
\def\contentsname{\small \'Indice}

\def\chaptername{}

\def\figurename{\small Fig.}

\usepackage{fancyhdr}
\usepackage{sectsty} 
\allsectionsfont{\raggedright} 
\makeindex

\begin{document}

\renewcommand{\figurename}{\textbf{Figura}}

\def\chaptername{}
\def\contentsname{Sum\'ario}
\pagenumbering{roman}


\parskip 3mm


\pagestyle{plain}
\thispagestyle{empty}
\pagenumbering{arabic}
\parskip 3mm
\baselineskip 13.1pt

\begin{flushleft}
\begin{minipage}{13.cm}
\baselineskip=24pt
\chapter*{Giotto e Galileu: novos olhares sobre o C\'{e}u e sobre o Livro da Natureza}

\vspace*{0.5cm}
\noindent{\Large\textbf{Francisco Caruso}}
\end{minipage}
\end{flushleft}

\vspace*{1.0cm}

\noindent {\textbf{Resumo:} Neste trabalho, apresenta-se uma tentativa (incompleta) de compre\-en\-der a con\-tri\-bui\-\c{c}\~{a}o de Galileu Galilei no \^{a}mbito da Hist\'{o}ria das Ideias, a partir do momento em que aponta uma luneta para o c\'{e}u. Para isso, foi preciso, antes, que Giotto pintasse o c\'{e}u de azul. Reconstru\'{\i}mos toda essa fascinante trajet\'{o}ria entre os s\'{e}culos XIII e XVII na Europa, relacionando diferentes fazeres e saberes, bem como discutindo o pensamento cient\'{\i}fico e filos\'{o}fico daquela \'{e}poca. A tese principal desse ensaio \'{e} mostrar que Giotto, quem pintou o c\'{e}u de azul numa cultura que o desejava dourado, percorreu uma trajet\'{o}ria conceitual muito semelhante \`{a}quela trilhada por Galileu s\'{e}culos depois.

\newpage
%

\section{Introdu\c{c}\~{a}o}

Tudo come\c{c}ou quando o f\'{\i}sico pisano Galileu Galilei (1564-1642), pela primeira vez na hist\'{o}ria (1609), apontou uma luneta para o c\'{e}u (Figuras~\ref{fig-caruso-1} e \ref{fig-caruso-2}), numa ati\-tu\-de manifestamente questionadora, desafiadora, com um claro objetivo de fazer observa\c{c}\~{o}es astron\^{o}micas de cunho cient\'{\i}fico. Assim, fez descobertas im\-por\-tantes: por exemplo, que J\'{u}piter possui sat\'{e}lites, que a superf\'{\i}cie do Sol tem manchas e a Lua, irregularidades representadas por montanhas e crateras, marcando o fim de um longo per\'{\i}odo de observa\c{c}\~{o}es celestes a olho nu~\cite{Danhoni} e com\-pro\-me\-ten\-do a cren\c{c}a, ent\~{a}o vigente, na perfei\c{c}\~{a}o do c\'{e}u, incluindo os corpos celestes. Co\-me\-\c{c}aram, nessa hora, seus problemas com a Igreja, que, naquela \'{e}poca, ainda exercia enorme influ\^{e}ncia at\'{e} mesmo sobre a Ci\^{e}ncia. Em meio a essa crise, floresceu uma Nova Astronomia. A bem da verdade, se estava testemunhando a fase embrion\'{a}ria de um novo m\'{e}todo cient\'{\i}fico~\cite{Finocchiaro,Mariconda}. Este ensaio \'{e} dedicado a esse gesto -- \textit{tutt'altro che semplice} -- de Galileu, fruto da vontade e determina\c{c}\~{a}o de lan\c{c}ar um novo olhar sobre o C\'{e}u. Procura-se aqui compreender o pano de fundo cultural e cien\-t\'{\i}\-fico que propiciou tal gesto, incluindo algumas de suas premissas~\cite{Wallace} e con\-se\-qu\^{e}ncias~\cite{Geymonat}.

Quanto \`{a} serventia da luneta, o f\'{\i}sico brasileiro e caro amigo Roberto Moreira Xavier de Ara\'{u}jo disse uma vez: ``Uma sociedade que pinta o c\'{e}u de dourado n\~{a}o pode apontar uma luneta para o c\'{e}u''~\cite{frase}. Verdade, pois o c\'{e}u dourado \'{e} uma express\~{a}o inequ\'{\i}voca do car\'{a}ter divino a ele atribu\'{\i}do, lugar sagrado da morada de Deus no ima\-gi\-n\'{a}rio crist\~{a}o, algo a ser, portanto, contemplado, admirado, respeitado. Con\-se\-quentemente, n\~{a}o \'{e} pass\'{\i}vel de ser questionado, escrutinado e tampouco objeto de investiga\c{c}\~{a}o cient\'{\i}fica be\-ne\-fi\-ciando-se de uma luneta~\cite{Dourado}.

\begin{figure}[hbtp]
\centerline{\includegraphics[width=6.1cm]{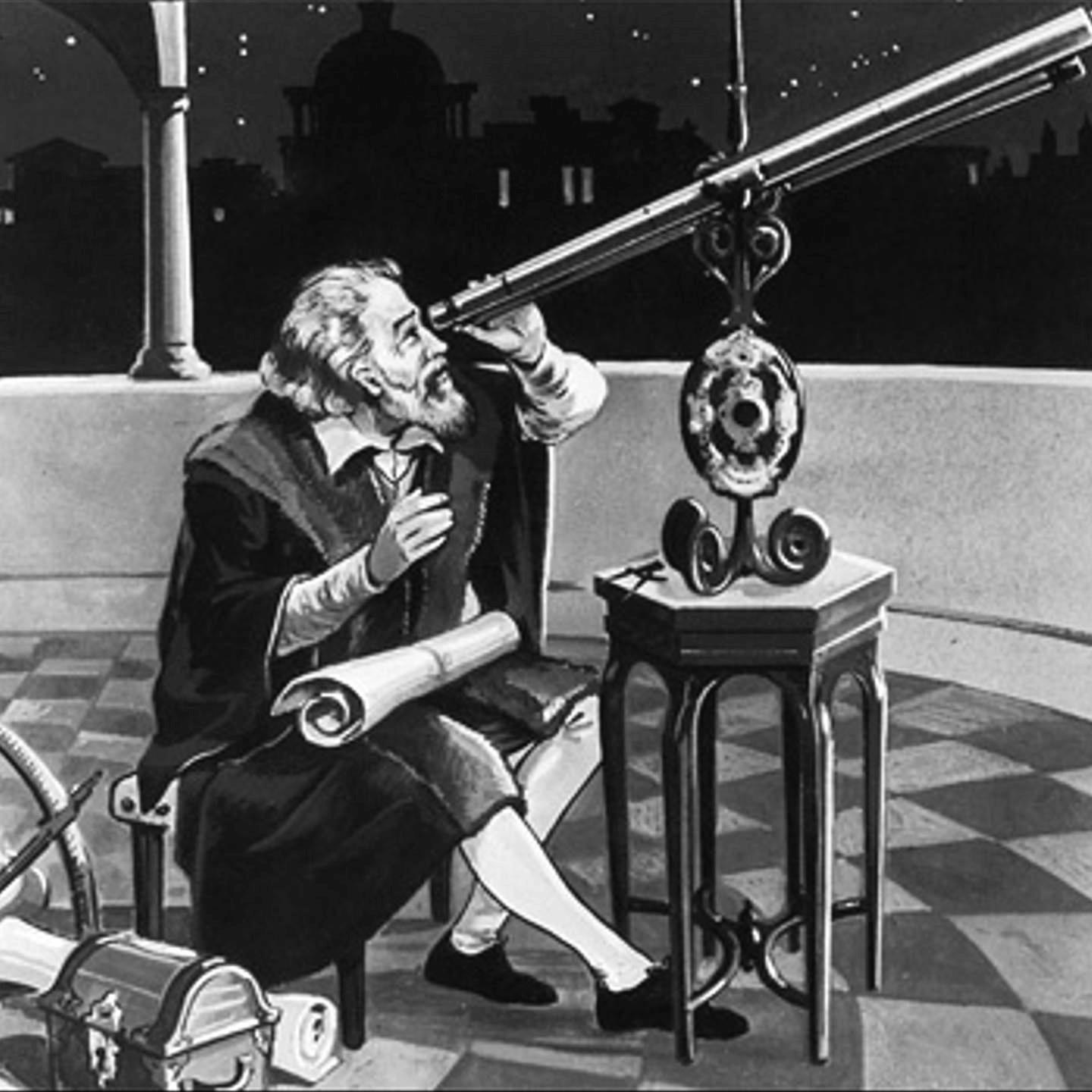}}
\caption{\small{Ilustra\c{c}\~{a}o de Galileu apontando uma luneta para o c\'{e}u. Cr\'{e}dito: Wikimedia Commons.}}
\label{fig-caruso-1}
\end{figure}

\begin{figure}[htbp]
\centerline{\includegraphics[width=13cm]{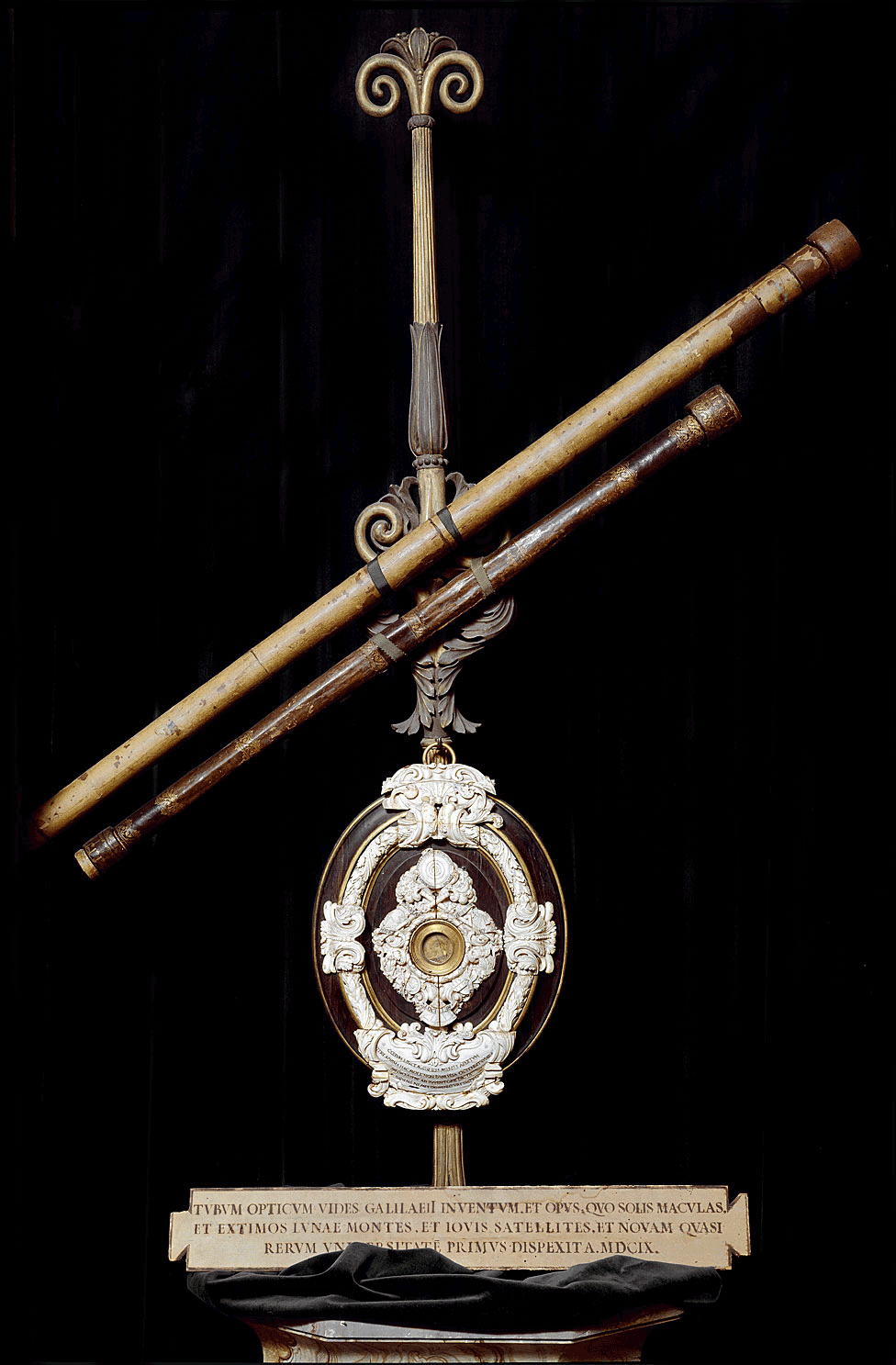}}
\caption{\small{Lente objetiva e luneta astron\^{o}mica da Galileia ou Galileu Galilei, 1610. \textit{Museo di Storia della Scienza}, Floren\c{c}a.}}
\label{fig-caruso-2}
\end{figure}

\newpage

Deixar de pintar o c\'{e}u de dourado \'{e} uma atitude que se insere num contexto bem mais amplo do que uma simples decis\~{a}o voluntariosa; na verdade, esse ato \'{e} fruto de um importante momento hist\'{o}rico de transi\c{c}\~{a}o no Mundo das Ideias e das Artes,\footnote{\, A Ci\^{e}ncia n\~{a}o ficar\'{a} imune a essas mudan\c{c}as no modo de ver e representar o Mundo.} no qual se destaca a valoriza\c{c}\~{a}o da Natureza como \textit{ela \'{e} de fato}. Nesse sentido, pelo pioneirismo, dois nomes se sobressaem: S\~{a}o Francisco de Assis e o pintor e arquiteto florentino Giotto di Bondone. Suas contribui\c{c}\~{o}es ser\~{a}o apresentadas e analisadas na pr\'{o}xima Se\c{c}\~{a}o, pois, em nossa opini\~{a}o, no plano da Hist\'{o}ria das Ideias, s\~{a}o relevantes para a prepara\c{c}\~{a}o de um solo cultural f\'{e}rtil no qual Galileu pudesse efetivamente direcionar sua luneta para o C\'{e}u azul.

\section{Mudan\c{c}as na Arte no s\'{e}culo XIII: do c\'{e}u dourado ao c\'{e}u azul}\label{Arte}

Era uma tradi\c{c}\~{a}o na Arte Bizantina que o c\'{e}u fosse pintado de dourado~\cite{Mango}, como nessa obra cl\'{a}ssica de Duccio~(Fig.~\ref{fig-caruso-3}).

\begin{figure}[hbtp]
\centerline{\includegraphics[width=9.3cm]{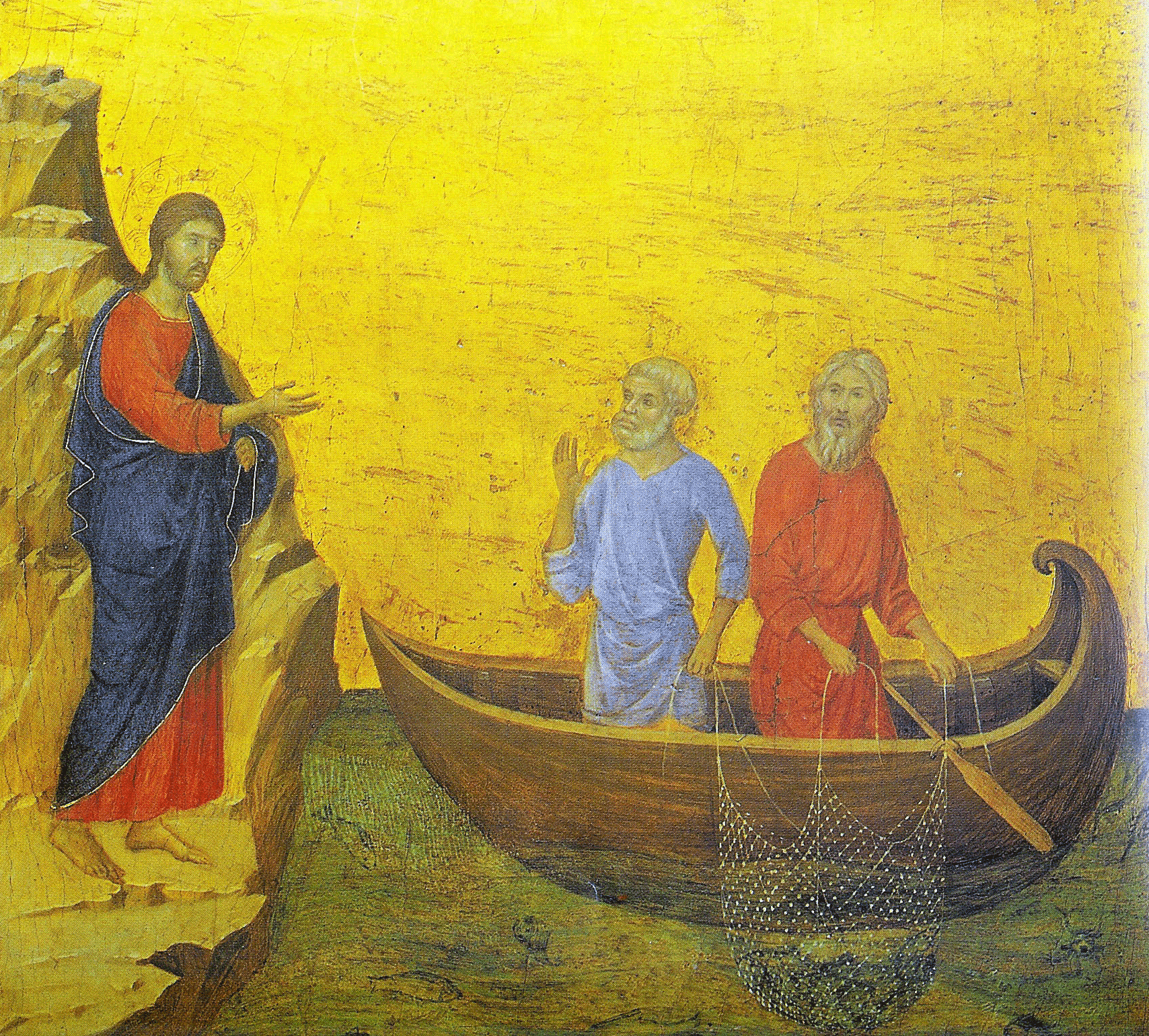}}
\caption{\small{``O Chamado dos Ap\'{o}stolos Pedro e Andr\'{e}'', de Duccio Buoninsegna, c.~1308-1311. National Gallery of Art, Washington. Trata-se de um painel que chegou a fazer parte da predela da obra \textit{Maest\`{a}}, considerada a maior cria\c{c}\~{a}o individual da Escola Bizantina, embora tenha sido pintada por Duccio, entre 1308 e 1311, e uma das mais perfeitas express\~{o}es da Arte Medieval.}}
\label{fig-caruso-3}
\end{figure}

Esta particularidade recorrente de representa\c{c}\~{a}o do c\'{e}u tem ra\'{\i}zes no fato de a Arte Medieval ter se desenvolvido durante um longo per\'{\i}odo no qual a Igre\-ja Cat\'{o}lica, extrapolando assuntos religiosos, influenciava, supervisionava e fil\-tra\-va todas as produ\c{c}\~{o}es culturais e mesmo as cient\'{\i}ficas. A Arte desse per\'{\i}odo serviu, em boa parte, para a decora\c{c}\~{a}o de igrejas, catedrais, mosteiros e pal\'{a}cios. Portanto, essencialmente, seu papel durante um longo per\'{\i}odo foi o de aproximar o Homem de Deus, as pessoas da religiosidade, reafirmando a sociedade e o estado teoc\^{e}ntricos caracter\'{\i}sticos da Idade M\'{e}dia. O abandono desse c\'{e}u dourado, tra\-ta\-do a seguir, \'{e} um press\'{a}gio de uma importante transforma\c{c}\~{a}o na vis\~{a}o de mundo, que come\c{c}a a mudar com S\~{a}o Francisco de Assis~\cite{Vitruvian}.

A relev\^{a}ncia da contribui\c{c}\~{a}o franciscana para a forma\c{c}\~{a}o de um novo homem pode ser aferida pelas palavras do escritor brit\^{a}nico Gilbert Keith Chesterton~\cite{Chesterton} quando ele afirma que o aco\-lhi\-mento do pensamento de S\~{a}o Francisco
\begin{quotation}
\noindent \small{marcou o momento em que os homens puderam se reconciliar n\~{a}o apenas com Deus, mas com a natureza e, o mais dif\'{\i}cil de tudo, consigo mesmos.}
\end{quotation}
\normalsize

Do ponto de vista do tema que estamos abordando, cabe enfatizar que S\~{a}o Fran\-cisco contribuiu, efetivamente, para a difus\~{a}o da met\'{a}fora do \textit{Livro da Natureza}\footnote{\, A met\'{a}fora do \textit{Livro da Natureza} j\'{a} era conhecida desde os prim\'{o}rdios do cristianismo. Ela ganha for\c{c}a no final do s\'{e}culo~IV, no pensamento de Santo Agostinho. De acordo com ele, h\'{a} dois livros que podem nos levar ao conhecimento de Deus: o \textit{Livro da Natureza} e o da \textit{Sagrada Escritura}. Essa cren\c{c}a de que Deus se revela em ambos influenciou fortemente a percep\c{c}\~{a}o crist\~{a} do mundo durante a Idade M\'{e}dia. Mas \'{e} a partir de S\~{a}o Francisco de Assis que essa ideia vai se difundir mais e dar suporte \`{a}s mudan\c{c}as no pr\'{o}prio pensamento filos\'{o}fico-cient\'{\i}fico. De todo modo, considere-se tamb\'{e}m a engenhosidade e o alcance desse argumento tendo em vista que, naquela \'{e}poca, uma enorme parcela da popula\c{c}\~{a}o era analfabeta e, dessa forma, n\~{a}o poderia ler a B\'{\i}blia.} como caminho alternativo \`{a} B\'{\i}blia para o homem que, contemplando a maravilha da Cria\c{c}\~{a}o, chegaria igualmente at\'{e} Deus~\cite{livro-espaco-natureza}. Tal met\'{a}fora vai ser empregada, mais tarde, por Dante Alighieri, na \textit{Divina Com\'{e}dia}, e pelo pr\'{o}prio Galileu, em \textit{O En\-sa\'{\i}sta}, dentre outros autores.

Em ess\^{e}ncia, de uma perspectiva filos\'{o}fica, ao defender essa nova via para se chegar a Deus, S\~{a}o Francisco sugere que se olhe para a Natureza \textit{como ela \'{e}}. Dessa forma, ele est\'{a} antecipando em muito uma das caracter\'{\i}sticas mais not\'{a}veis do \textit{Renascimento}: um olhar desarmado, mais livre, mais objetivo e cr\'{\i}tico, menos carregado de simbolismos (t\~{a}o em voga na Idade M\'{e}dia), que, em \'{u}ltima an\'{a}lise, contribuir\'{a} mais tarde para a forma\c{c}\~{a}o de um novo sujeito, como aponta o fil\'{o}sofo alem\~{a}o Ernst Cassirer~\cite{Cassirer}.

A partir de S\~{a}o Francisco de Assis, o peso relativo do \textit{Livro da Natureza} como ins\-trumento capaz de levar a Deus -- ou por meio do qual Ele oferece salva\c{c}\~{a}o ao Homem -- come\c{c}a a mudar, reflexo de um novo olhar sobre a \textit{Natureza}, um olhar, digamos, mais realista. Nessa nova \'{o}ptica, S\~{a}o Francisco sugere que se busque, na simplicidade e na harmonia das coisas, a beleza suprema da obra divina.
Assim, admirar a beleza do mundo passa a ser um caminho alternativo (e n\~{a}o menos digno) de se chegar a Deus. Dessa forma, ele deu os primeiros passos para uma nova humaniza\c{c}\~{a}o do Mundo, admitindo que o homem deve se integrar \`{a} Natureza, sem se afastar de Deus, preservando-a e admirando-a na sua plenitude.

Admirador de Francisco e sob o impacto de seus ensinamentos, Giotto ser\'{a} o pri\-mei\-ro pintor a representar os santos com apar\^{e}ncias humanas e o c\'{e}u n\~{a}o mais dou\-ra\-do -- caracter\'{\i}stico da arte bizantina --, mas azul (Fig.~\ref{fig-caruso-4})~\cite{Dourado}. Ser\'{a} tamb\'{e}m o pio\-nei\-ro na introdu\c{c}\~{a}o do espa\c{c}o tridimensional na pintura italiana por meio da perspectiva~\cite{Edgerton}. Esse conjunto de atitudes no mundo das Artes \'{e} essencial para a forma\c{c}\~{a}o de um novo \textit{Zeitgeist}. Se, como defendia o fil\'{o}sofo alem\~{a}o Georg Wilhelm Friedrich Hegel, por sua pr\'{o}pria natureza, a arte reflete a cultura da \'{e}poca (representada aqui pelo pensamento franciscano) na qual est\'{a} inserida, sob outra perspectiva, ter\'{a} impactos sociais, culturais e at\'{e} mesmo cient\'{\i}ficos com o passar do tempo -- como efetivamente teve --, alargando esse ``esp\'{\i}rito da \'{e}poca'', o que ser\'{a} discutido ao longo deste ensaio.

\begin{figure}[hbtp]
\centerline{\includegraphics[width=10.8cm]{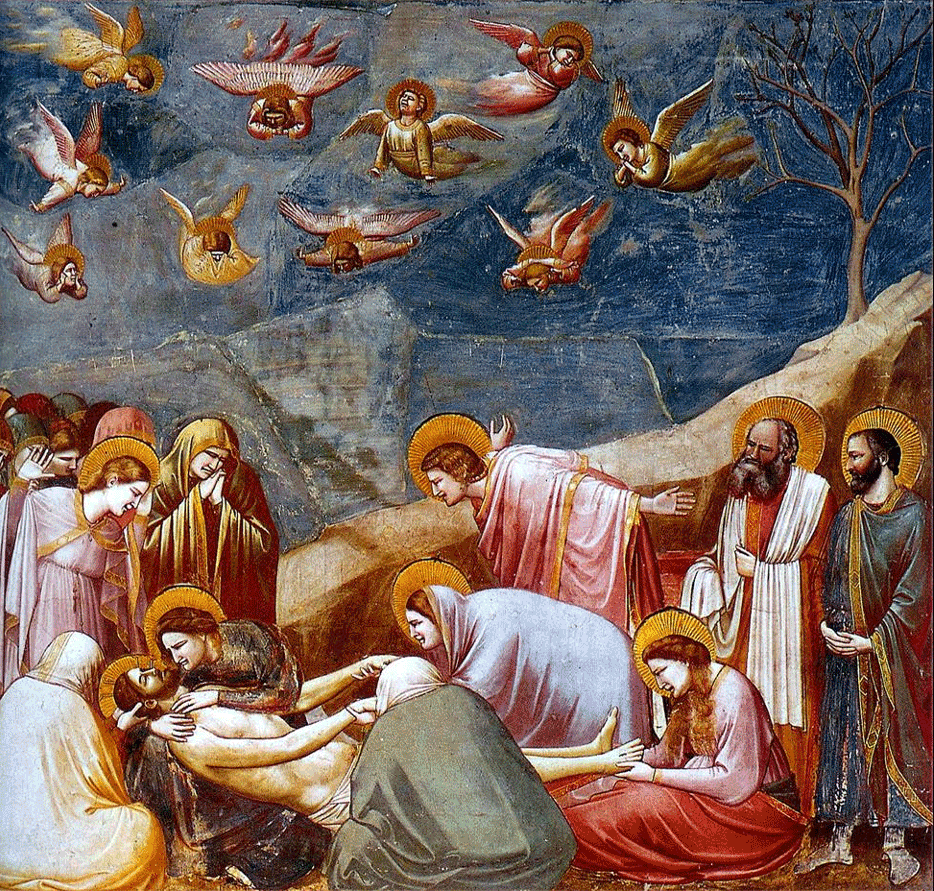}}
\caption{\small{``A Lamenta\c{c}\~{a}o'', c.~1305, afresco pintado por Giotto, como parte de seu ciclo da Vida de Cristo na parede interna norte da Capela Scrovegni em P\'{a}dua, It\'{a}lia.}}
\label{fig-caruso-4}
\end{figure}
\newpage

Para Giotto, Francisco \'{e} um ser humano, semelhante aos demais, e sua lenda torna-se hist\'{o}ria, impregnada da realidade do cotidiano e da concretude de uma cidade, de uma paisagem, de um mundo real conhecido. Trata-se de uma re\-vo\-lu\-\c{c}\~{a}o figurativa, pois d\'{a} in\'{\i}cio a um processo de seculariza\c{c}\~{a}o do sagrado e de atualiza\c{c}\~{a}o dos acontecimentos absolutamente distinto -- na verdade oposto -- da tradi\c{c}\~{a}o medieval, e que caracterizar\'{a} a pintura, em particular, ao longo do s\'{e}culo XIV e posteriormente. Logo, a representa\c{c}\~{a}o da arte sacra nas paredes da Bas\'{\i}lica Superior de Assis, onde Giotto pintou a vida de S\~{a}o Francisco,\footnote{\, Giotto pintou quase todas as paredes da Igreja Inferior de S\~{a}o Francisco, em Assis. ``A Lenda de S\~{a}o Francisco'', cuja autoria \'{e} tradicionalmente atribu\'{\i}da tamb\'{e}m a ele, \'{e} o tema de um ciclo de 28 afrescos que se encontram na Igreja Superior de S\~{a}o Francisco, pintados, provavelmente, entre 1297 e 1300.} carrega uma conota\c{c}\~{a}o bem diferente: secular, profana, moderna.

Tais feitos s\~{a}o, em suma, reflexos de uma nova dupla atitude de inspira\c{c}\~{a}o franciscana: observar di\-re\-tamente a Natureza (que o leva, por exemplo, a pintar o c\'{e}u de azul), e tentar imitar o mundo \textit{natural}, o que traduz um interesse real no mundo \textit{material} (como santos sendo pintados inspirando-se em moradores da cidade) e n\~{a}o apenas no mundo \textit{espiritual}~\cite{Bunim}. Ambas as atitudes requerem uma vis\~{a}o atenta e determinada, e a capacidade de desvendar as propor\c{c}\~{o}es naturais daquilo que se v\^{e}, o que, em \'{u}ltima an\'{a}lise, remete ao estudo da Geometria e da \'{O}ptica~\cite{Optica}. Vale lembrar que, na Idade M\'{e}dia, os termos \textit{\'{o}ptica} e \textit{perspectiva} eram, muitas vezes, usados indiscriminadamente. Cabe ainda enfatizar que, embora incipiente em seu come\c{c}o, trata-se de uma tend\^{e}ncia duradoura que aponta para uma aproxima\c{c}\~{a}o entre Arte e Ci\^{e}ncia, cujo \'{a}pice ocorrer\'{a} no Renascimento~\cite{White}.

Nesse aspecto, em resumo, tr\^{e}s s\~{a}o os pontos fundamentais dessa nova dis\-po\-si\-\c{c}\~{a}o na Arte: a tentativa de sua geometriza\c{c}\~{a}o, a busca de sua seculariza\c{c}\~{a}o e a procura de representar o movimento. Todos esses pontos estar\~{a}o mais tarde envolvidos, de algum modo, na constru\c{c}\~{a}o de uma Nova Ci\^{e}ncia Natural.

O historiador e te\'{o}rico italiano da Arte Giulio Carlo Argan nos ensina que, du\-ran\-te todo o curso do s\'{e}culo XIII, a Arte Italiana, em especial a pintura, passa por um processo de abandono de uma mera apresenta\c{c}\~{a}o da imagem para a representa\c{c}\~{a}o de a\c{c}\~{o}es~\cite{Argan}. Dessa tend\^{e}ncia, afirma Argan, a ``ideologia bizantina do eterno'' passa a ser substitu\'{\i}da pela ``ideologia da hist\'{o}ria'' e, dessa forma, contribui para formar, em suas palavras, ``o pensamento de que a consci\^{e}ncia da hist\'{o}ria seja a base de todo interesse cognitivo e \'{e}tico''.

Ernst Gombrich, outro historiador da Arte, de origem austr\'{\i}aca, sustenta que uma boa maneira de aferir a magnitude da revolu\c{c}\~{a}o feita por Giotto na Arte \'{e} comparar a pintura da Fig.~\ref{fig-caruso-4} com outra cena do sepultamento de Cristo, tamb\'{e}m do s\'{e}culo XIII, descrita em uma miniatura encontrada em um salt\'{e}rio manuscrito. Essa sua an\'{a}lise esclarecedora \'{e} apresentada em seu livro~\cite{Gombrich} e vale a pena ser consultada. Giotto redescobre a arte de criar a ilus\~{a}o de profundidade numa su\-per\-f\'{\i}cie plana. Para isso, recorre \`{a} Perspectiva, mas n\~{a}o como a conhecemos hoje. A pers\-pectiva por ele adotada foi chamada de ``espinha de peixe'', porque as linhas de fuga n\~{a}o convergiam para um \'{u}nico ponto, mas para v\'{a}rios pontos dispostos ao logo de um eixo. Mas, de qualquer maneira, foi desse jeito que Giotto pode su\-pe\-rar a bidimensionalidade da arte bizantina e obter um maior rea\-lis\-mo na re\-pre\-sen\-ta\c{c}\~{a}o das suas cenas e figuras.

Ainda segundo Gombrich, foi dessa maneira que ele quebrou o sortil\'{e}gio do con\-ser\-va\-do\-rismo da arte bizantina, ``aventurando-se em um mundo, no qual tra\-duz para a pintura figuras realistas da escultura g\'{o}tica'', a partir de uma mo\-de\-la\-\c{c}\~{a}o de luz e sombra~\cite{Baxandall} e, n\~{a}o \'{e} demais acrescentar, recorrendo \`{a} Geo\-metria.

Em suma, a partir do s\'{e}culo XIII, identificamos algumas mudan\c{c}as conceituais e novas tend\^{e}ncias t\'{e}cnicas que apontam para uma revolu\c{c}\~{a}o na Arte, que se ins\-taura aqui e culmina no Renascimento. S\~{a}o elas:
\begin{enumerate}
  \item valoriza\c{c}\~{a}o do indiv\'{\i}duo e da hist\'{o}ria;
  \item tentativa de seculariza\c{c}\~{a}o das representa\c{c}\~{o}es pict\'{o}ricas;
  \item busca por uma representa\c{c}\~{a}o mais real da Natureza;
  \item incorpora\c{c}\~{a}o da Geometria por meio da perspectiva;
  \item considera\c{c}\~{a}o do movimento e sua poss\'{\i}vel representa\c{c}\~{a}o.
\end{enumerate}

Todas essas caracter\'{\i}sticas estar\~{a}o igualmente presentes no vindouro embate em torno do fazer cient\'{\i}fico (Se\c{c}\~{a}o~\ref{Viri}), com ineg\'{a}vel impacto epistemol\'{o}gico e metodol\'{o}gico sobre a \textit{Ci\^{e}ncia Moderna}~\cite{Jenner,Leituras}.

Entretanto, antes de abordar a contribui\c{c}\~{a}o de Galileu \`{a} Astronomia e \`{a} F\'{\i}sica, \'{e} importante discutir, ainda que resumidamente, alguns t\'{o}picos: o papel da \'{O}ptica no s\'{e}culo~XIII (Se\c{c}\~{a}o~\ref{Optica}); em seguida, a relev\^{a}ncia da Geometria tamb\'{e}m na As\-tro\-nomia (Se\c{c}\~{a}o~\ref{Geo}) e, por \'{u}ltimo, apresentar uma brev\'{\i}ssima hist\'{o}ria da Abo\-li\c{c}\~{a}o do Cosmos Aristot\'{e}lico (Se\c{c}\~{a}o~\ref{Cosmos}) que sucedeu ao copernicanismo e est\'{a} no centro da pol\^{e}mica na qual Galilei se viu envolvido com colegas conservadores e com a Igreja.

\section{O papel da \'{O}ptica na Arte e na Ci\^{e}ncia no s\'{e}culo XIII}\label{Optica}

Euclides apresentou a primeira teoria matem\'{a}tica da vis\~{a}o, na melhor tradi\c{c}\~{a}o pla\-t\^{o}\-nica, concentrando-se na fundamenta\c{c}\~{a}o geom\'{e}trica da \textit{\'{O}ptica}. Ignorando a causa primeira dos fe\-n\^{o}\-me\-nos e qualquer interesse fisiol\'{o}gico, preocupou-se apenas com o que \'{e} observado e pode ser expresso geometricamente, seguindo o m\'{e}todo detalhado em seus \textit{Elementos}. Sua contribui\c{c}\~{a}o ao ideal de geometrizar a F\'{\i}sica, em certo sentido, foi al\'{e}m de Plat\~{a}o -- que se restringiu a estabelecer as bases de uma Cosmologia e de uma vis\~{a}o da estrutura da mat\'{e}ria fundamentadas no Mundo das Ideias --, no \^{a}mbito do que se pode chamar de uma filosofia geom\'{e}trica. De fato, Euclides lan\c{c}ou m\~{a}o de uma descri\c{c}\~{a}o geom\'{e}trica dos fe\-n\^{o}\-menos luminosos, no sentido mais atual desse termo -- um sentido quase galileano --, separando a quest\~{a}o da propaga\c{c}\~{a}o f\'{\i}sica da luz de outras para as quais ele n\~{a}o teria resposta. Ao aplicar a Geometria ao estudo da \'{O}ptica, mostrou que fen\^{o}menos reais podiam ser descritos qualitativa e quantitativamente. Esse foi um marco importante, normalmente n\~{a}o enfatizado, na hist\'{o}ria da geometriza\c{c}\~{a}o da F\'{\i}sica, que contar\'{a}, mais tarde, com as contribui\c{c}\~{o}es de Ren\'{e} Descartes, Galileu e Isaac Newton, e ganhar\'{a} uma nova dimens\~{a}o com Albert Einstein~\cite{Caruso-Oguri}.

No s\'{e}culo XIII, como fruto da postura franciscana com rela\c{c}\~{a}o \`{a} Natureza, e n\~{a}o por acaso, muitos religiosos dessa Ordem dedicaram-se ao estudo da \textit{\'{O}ptica} numa \'{e}poca em que a Ci\^{e}ncia estava sob o controle da Igreja.

\'{E} importante recordar que a luz \'{e} algo que sempre fascinou o homem. O brilho das estrelas, o Sol, o fogo. Aquilo que o afasta da escurid\~{a}o que o amedronta. Aquilo que transita entre o C\'{e}u e a Terra, merecendo, por isso, um lugar de des\-ta\-que em seu imagin\'{a}rio. Umberto Eco chama a aten\c{c}\~{a}o para o fato de a luz ser \'{u}nica quando se considera a possibilidade de transitar entre o C\'{e}u e a Terra, que Deus separou~\cite{Eco}. Essa caracter\'{\i}stica, al\'{e}m de obviamente fomentar o ima\-gi\-n\'{a}rio coletivo religioso, acaba tendo um papel importante no estudo da \'{O}ptica, no \textit{Medioevo}, ao mesmo tempo em que \'{e} ressaltada na Arte~\cite{Vitruvian}.

Ousamos afirmar que, naquela sociedade medieval, os v\'{a}rios interesses en\-vol\-ven\-do o estudo da luz acabaram aproximando Arte e Ci\^{e}ncia tendo a Geo\-me\-tria como linguagem comum, em um momento hist\'{o}rico \'{u}nico. Efe\-ti\-va\-men\-te, no cerne de uma sociedade ainda eminentemente teoc\^{e}ntrica, interesses cien\-t\'{\i}\-fi\-cos, filos\'{o}ficos e art\'{\i}sticos se avizinham, se entrela\c{c}am e, de certa forma, se con\-fundem. N\~{a}o identificamos nenhuma outra \'{e}poca em que tal fato seja reincidente.

A partir desse per\'{\i}odo, h\'{a} uma prolifera\c{c}\~{a}o de livros e tratados sobre Perspectiva e a utiliza\c{c}\~{a}o da Geometria na Pintura, muitos escritos por artistas famosos, como foi revisto em~\cite{Vitruvian}. Vamos, aqui, nos ater apenas a mais alguns coment\'{a}rios relacionados \`{a} Ci\^{e}ncia, objetivo principal deste ensaio.

De volta \`{a} contribui\c{c}\~{a}o cient\'{\i}fica dos franciscanos, sabe-se que o bispo, te\'{o}logo e erudito ingl\^{e}s Robert Grosseteste, fundador da Escola Franciscana de Oxford, muito estimulou o interesse pela \'{O}ptica na Europa, no s\'{e}culo XIII. Ele considerava, por exemplo, que a luz, por sua extens\~{a}o, condensa\c{c}\~{a}o e rarefa\c{c}\~{a}o, explicava todos os fen\^{o}menos do Universo. Roger Bacon, por volta de 1240, ingressou para a Ordem Franciscana, onde, sob influ\^{e}ncia de Grosseteste, dedicou-se a estudos da perspectiva e outros temas coligados,\footnote{\,No que se refere especificamente \`{a} \'{O}ptica, tanto Grosseteste quanto Bacon tiveram enorme influ\^{e}ncia do pol\'{\i}mata persa Ibn Al-Haytham -- conhecido no Ocidente como Alhazen -- que havia escrito o importante livro \textit{De aspectibus}, com uma forte descri\c{c}\~{a}o matem\'{a}tica da \'{O}ptica, explorando sua natureza geom\'{e}trica.} introduzindo a observa\c{c}\~{a}o da natureza e a experimenta\c{c}\~{a}o como fundamentos do conhecimento natural. Repete-se aqui a tend\^{e}ncia de valorizar o mundo natural. Na verdade, Bacon foi al\'{e}m de seu tutor, antecipando concep\c{c}\~{o}es de Galileu que revolucionaram a F\'{\i}sica e marcaram o que se passou a denominar \textit{Ci\^{e}ncia Moderna}, ao afirmar que o m\'{e}todo cient\'{\i}fico depende da observa\c{c}\~{a}o, da experimenta\c{c}\~{a}o, da elabora\c{c}\~{a}o de hip\'{o}teses e da ne\-ces\-sidade de verifica\c{c}\~{a}o independente.

Quanto a essa compreens\~{a}o embrion\'{a}ria de um novo m\'{e}todo cient\'{\i}fico, es\-bo\-\c{c}ada no final da Idade M\'{e}dia e que acaba de ser descrita muito es\-que\-ma\-ti\-ca\-mente, \'{e} ela que, mais tarde, em \'{u}ltima an\'{a}lise, ir\'{a} libertar de vez a Ci\^{e}ncia e, em par\-ti\-cular, a Astronomia, de todo um conjunto de atitudes cerceadoras impostas, me\-ta\-foricamente, pela representa\c{c}\~{a}o sistem\'{a}tica do c\'{e}u dourado. Nesse processo, o astr\^{o}nomo polon\^{e}s Nicolau Cop\'{e}rnico desempenhar\'{a} um papel fundamental, como ser\'{a} relembrado na pr\'{o}xima Se\c{c}\~{a}o.

Por fim, vale a pena comentar que o interesse inicial de Galileu pelo telesc\'{o}pio limitava-se \`{a} pr\'{o}pria \'{O}ptica. Ele teve que aprender a fazer lentes e, com o fruto do seu trabalho, conseguiu aumentar a amplia\c{c}\~{a}o do telesc\'{o}pio original em cerca de 10 vezes, o que, na pr\'{a}tica, viabilizou suas observa\c{c}\~{o}es.

\section{A relev\^{a}ncia da Geometria para a Arte e a Astronomia}\label{Geo}

Embora as formas geom\'{e}tricas simples tenham contribu\'{\i}do para a elabora\c{c}\~{a}o de arqu\'{e}tipos medievais~\cite{Zumthor}, devemos destacar o crescente in\-te\-resse pela Geo\-me\-tria nos \'{u}ltimos s\'{e}culos da Idade M\'{e}dia, associado, em um primeiro momento, ao soerguimento da Tecnologia entre os s\'{e}culos XI e XIV, notadamente: a in\-ven\-\c{c}\~{a}o do moinho de vento e de artefatos mec\^{a}nicos movidos por for\c{c}a hi\-dr\'{a}u\-li\-ca, aperfei\c{c}oamentos n\'{a}uticos e t\^{e}xteis, a inven\c{c}\~{a}o do rel\'{o}gio mec\^{a}nico e a cons\-tru\c{c}\~{a}o das catedrais.

No s\'{e}culo XIII, houve um grande interesse pelo estudo da \'{O}ptica, como j\'{a} men\-cionado, com impacto quase imediato nas Artes, via Geometria. De fato, ve\-ri\-ficou-se uma retomada de interesse por este ramo da Matem\'{a}tica no meio art\'{\i}stico e aca\-d\^{e}\-mico a partir desse s\'{e}culo, tanto na pintura at\'{e} o Renascimento Italiano, quan\-to, em uma segunda fase, na Astronomia, por inter\-m\'{e}dio do astr\^{o}nomo polon\^{e}s Nicolau Cop\'{e}rnico, do astr\^{o}nomo e matem\'{a}tico alem\~{a}o Johannes Kepler e do pr\'{o}prio Galileu~\cite{Cohen,Gingerich,Lattis}.

N\~{a}o \'{e} por acaso que, no s\'{e}culo XV, o livro \textit{Os Elementos da Geometria}, de Euclides, figurava entre as obras mais procuradas pelos estudiosos~\cite{McMurtrie}.

Trazemos, como \'{u}nico exemplo emblem\'{a}tico do emprego bem-sucedido da perspectiva na pintura,\footnote{\, Outros exemplos e uma discuss\~{a}o mais detalhada pode ser encontrada em~\cite{Vitruvian}.} o afresco \textit{Trinit\`{a}}, \'{u}ltima obra pintada por Masaccio, inspirada nos tra\-ba\-lhos do famoso arquiteto florentino Filippo Brunelleschi -- que havia definido a perspectiva como a ci\^{e}ncia da representa\c{c}\~{a}o. Esta obra \'{e} considerada pioneira quando se trata da correta utiliza\c{c}\~{a}o da perspectiva~(Fig.~\ref{fig-caruso-5}).

\begin{figure}[htbp]
\centerline{\includegraphics[width=7.0cm]{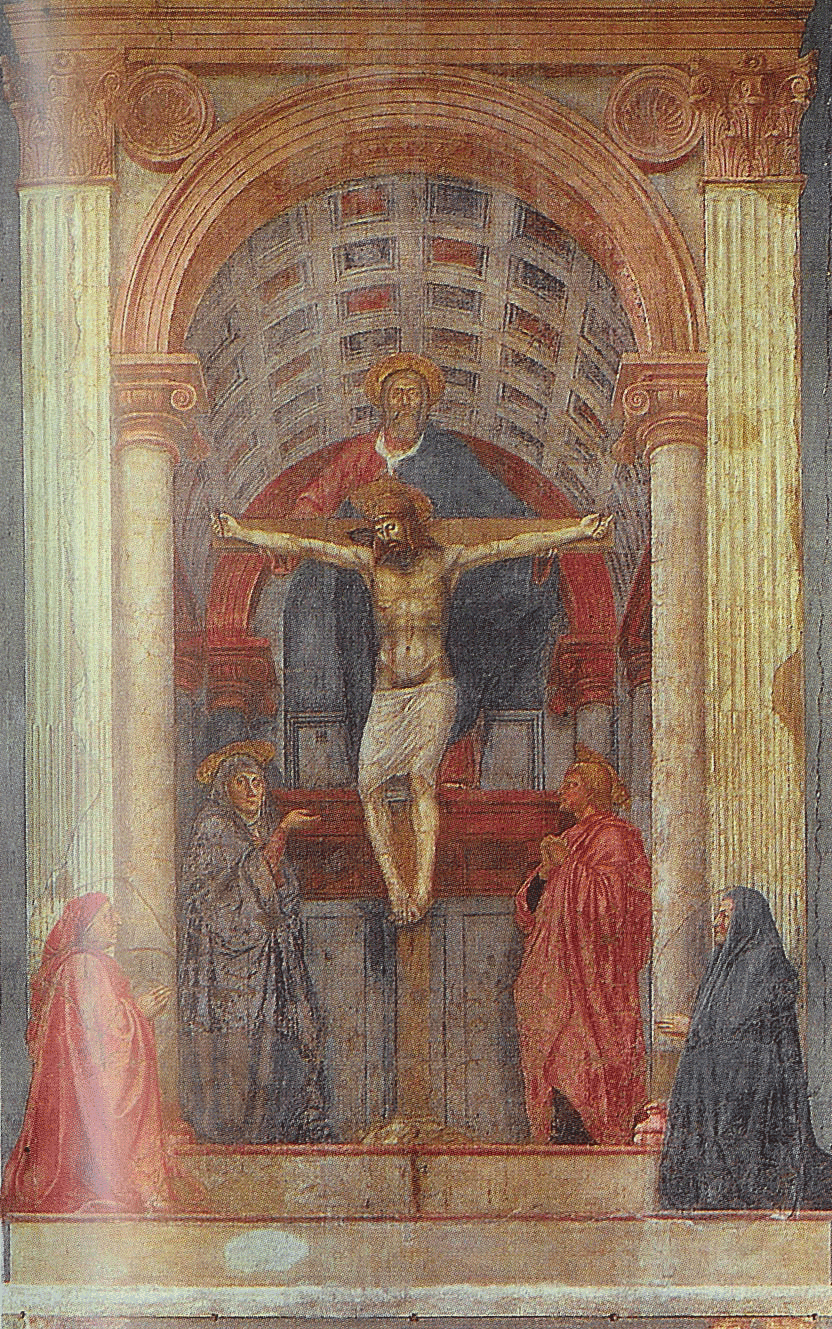}}
\caption{\small{``A Trindade'', afresco de Masaccio, 1426, encontra-se na Basilica di Santa Maria Novella, Floren\c{c}a, It\'{a}lia.}}
\label{fig-caruso-5}
\end{figure}

Passados pouco mais de 100 anos da consolida\c{c}\~{a}o da Perspectiva (geom\'{e}trica) na Pintura, nos deparamos com uma declara\c{c}\~{a}o expl\'{\i}cita de que tamb\'{e}m \'{e} a Geometria que vai nos levar a compreender e a representar os movimentos dos orbes celestes. Mais precisamente, estamos falando do ano de 1543, quando foi publicado postumamente o famoso tratado \textit{De Revolutionibus Orbium Coelestium} de Co\-p\'{e}r\-nico~\cite{Copernico}, que revolucionou o estudo dos c\'{e}us e abriu caminho para uma nova concep\c{c}\~{a}o de Mundo muito diferente do aristot\'{e}lico~\cite{Sobel}. Com efeito, o autor optou por imprimir na folha de rosto de seu livro uma frase em grego (assinalada em amarelo na Fig.~\ref{fig-caruso-6}) que, de certa forma, resume o enfoque de sua contribui\c{c}\~{a}o maior \`{a} Astronomia. \'{E} essencialmente a mesma frase que Plat\~{a}o teria mandado afixar na porta de sua Academia: ``\textit{Proibida a entrada a quem n\~{a}o conhece Geometria}''. Desta maneira, o astr\^{o}nomo polon\^{e}s deixa claro ao seu leitor, j\'{a} no frontisp\'{\i}cio de seu livro, que a chave para a compreens\~{a}o dos movimentos dos \textit{orbes} no C\'{e}u passa pela Geometria.

\begin{figure}[htbp]
\centerline{\includegraphics[width=6.5cm]{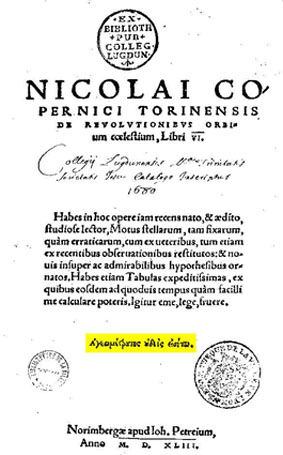}}
\caption{\small{Frontisp\'{\i}cio do livro de Cop\'{e}rnico, \textit{Sobre a revolu\c{c}\~{a}o dos Orbes Celestes}, publicado em 1543.}}
\label{fig-caruso-6}
\end{figure}

Passados pouco mais que 50 anos, no pr\'{o}dromo do livro \textit{Dissertatio cum Nuncio Sidereo accedit Narratio de quattro iovis satellitibus}, de Kepler~\cite{Kepler}, l\^{e}-se:

\begin{quotation}
\noindent \small{Disserta\c{c}\~{a}o cosmogr\'{a}fica contendo o mist\'{e}rio cosmogr\'{a}fico sobre as admi\-r\'{a}\-veis propor\c{c}\~{o}es dos orbes celestes, e sobre as raz\~{o}es pr\'{o}prias e ge\-nu\'{\i}\-nas do n\'{u}mero, da grandeza e dos movimentos peri\'{o}dicos dos c\'{e}us, demonstrado mediante os cinco corpos regulares da Geometria, por Johannes Kepler.}
\end{quotation}

O Kepler mais maduro vai abandonar esse modelo geom\'{e}trico plat\^{o}nico e a ideia aristot\'{e}lica das esferas celestes. Com base nos dados observacionais do astr\^{o}nomo dinamarqu\^{e}s Tycho Brahe, passa a buscar as leis matem\'{a}ticas das \textit{\'{o}rbitas} dos corpos celestes, expressas por curvas geom\'{e}tricas (elipses). Do ponto de vista metodol\'{o}gico, vemos, mais uma vez, uma mudan\c{c}a de atitude, novamente voltada para a valoriza\c{c}\~{a}o da observa\c{c}\~{a}o direta e sistem\'{a}tica do mundo natural.

Esse fato, em si, revela uma verdadeira revolu\c{c}\~{a}o intelectual, do ponto de vis\-ta das tentativas de se explicar fisicamente o movimento dos planetas. Sim, pois Kepler est\'{a} buscando, na verdade, uma unifica\c{c}\~{a}o da Astronomia com a F\'{\i}\-si\-ca, tentando estabelecer a causa din\^{a}mica das \'{o}rbitas dos planetas em torno do Sol, desafio esse que, contudo, s\'{o} foi resolvido por Newton~\cite{Westfall-vida}, com sua Teo\-ria da Gravita\c{c}\~{a}o Universal. S\'{o} destarte foi resolvida a crise de unidade da F\'{\i}\-si\-ca introduzida pelo helio\-cen\-trismo copernicano, contraposto ao geocentrismo aristot\'{e}lico.

\section{O heliocentrismo e a aboli\c{c}\~{a}o do Cosmos aristot\'{e}lico}\label{Cosmos}

Uma das maiores heran\c{c}as da hist\'{o}ria da humanidade \'{e} a constru\c{c}\~{a}o do que se pode chamar de cosmovis\~{a}o cient\'{\i}fica: um novo olhar sobre a Natureza, ou seja, sobre a \textit{Physis}, tal qual era entendida pelos gregos. A origem do processo de estrutura\c{c}\~{a}o dessa cosmovis\~{a}o, lento e fascinante, corresponde ao aparecimento e ao crescimento da Filosofia e da F\'{\i}sica na Gr\'{e}cia antiga. Fazem parte desse legado: a ruptura com explica\c{c}\~{o}es mitopo\'{e}ticas da natureza; o ideal de simplicidade manifestado desde quando se buscou compreender racionalmente a natureza a partir de um \'{u}nico princ\'{\i}pio, de uma mat\'{e}ria primordial organizada pela a\c{c}\~{a}o dos contr\'{a}rios (como queriam os pr\'{e}-socr\'{a}ticos); a procura de uma descri\c{c}\~{a}o da \textit{Physis} baseada em rela\c{c}\~{o}es causais, estabelecidas a partir da raz\~{a}o; e, finalmente, a concep\c{c}\~{a}o norteadora de que existe um \textit{Cosmos}, termo grego que significa um \textit{todo organizado}.

Apesar de marcantes diferen\c{c}as de pensamento, esse per\'{\i}odo cl\'{a}ssico da Fi\-lo\-sofia Grega caracteriza-se, em linhas gerais, pela presen\c{c}a do ideal de \textit{Cosmos} e pela convic\c{c}\~{a}o de que a ordena\c{c}\~{a}o da variedade
infinita das coisas e eventos possa (e deva) ser alcan\c{c}ada racionalmente. Isto posto, para os pensadores gregos, a compreens\~{a}o da Natureza passa necessariamente pela busca de um tipo de \textit{ordem}, o que, por
sua vez, requer o reconhecimento do que \'{e} igual, do que \'{e} regular ou, ainda, da capacidade de reconhecer simetrias: tudo em busca de uma \textit{Unidade}. Para os gregos, mesmo a beleza est\'{a} associada \`{a} ordem.

Arist\'{o}teles de Estagira defendia a exist\^{e}ncia de um Cosmos finito e ordenado no qual a Terra ocupava o centro. N\~{a}o era apenas o progresso ou a evolu\c{c}\~{a}o das coisas que ocorreria de forma ordenada; assim como Plat\~{a}o,
ele acreditava na exist\^{e}ncia de um \textit{telos} (de um \textit{fim}, uma \textit{finalidade}) ou, ainda, de uma perfei\c{c}\~{a}o, de uma ordem suprema, segundo a qual todas as transforma\c{c}\~{o}es acontecem por alguma ``viol\^{e}ncia'' ou para reparar a ordem previamente estabelecida do Cosmos provisoriamente rompida por um tipo de ``viol\^{e}ncia''. Essa hip\'{o}tese \'{e} conhecida como \textit{princ\'{\i}pio teleol\'{o}gico} e est\'{a} na base da filosofia aristot\'{e}lica.

Em resumo, o Cosmos aristot\'{e}lico \'{e} finito e hierarquizado, tendo a Terra no cen\-tro do Universo (Fig.~\ref{fig-caruso-7}), no qual n\~{a}o existe vazio~\cite{Koyre-zero}.

\begin{figure}[hbtp]
\centerline{\includegraphics[width=6.0cm]{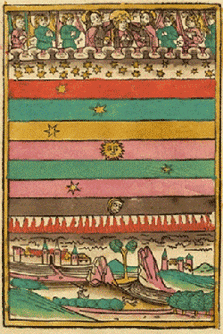}}
\caption{\small{Uma representa\c{c}\~{a}o medieval do Cosmos aristot\'{e}lico que originalmente ilustra o incun\'{a}bulo de autoria do escritor alem\~{a}o Konrad von Megenberg (1309-1374), \textit{Buch der Natur}, de 1481. Lessing
J.~Rosenwald Collection, Library of Congress, EUA.}}
\label{fig-caruso-7}
\end{figure}

Para o Estagirita, o Mundo sub-lunar \'{e} o mundo da corrup\c{c}\~{a}o, enquanto os c\'{e}us, tendo car\'{a}ter divino, s\~{a}o perfeitos e incorrupt\'{\i}veis. Em sua filosofia, C\'{e}u e Terra s\~{a}o ontologicamente diferentes.

No que concerne \`{a} Ci\^{e}ncia, \'{e} preciso ter em mente que o neoplatonismo e o neo-aristotelismo, por mais que tenham se afastado das doutrinas originais, foram essenciais para a descri\c{c}\~{a}o do mundo medieval, particularmente em torno do \textit{Timeu} de Plat\~{a}o e do Cosmos aristot\'{e}lico -- finito e hierarquizado~\cite{Koyre-zero}.
\newpage

O estudioso romeno das religi\~{o}es e da mitologia Mircea Eliade nos oferece uma jus\-ti\-fi\-cativa da ado\c{c}\~{a}o do Cosmos aristot\'{e}lico pela doutrina crist\~{a}~\cite{Eliade} que foi dis\-cu\-tida sucintamente em~\cite{livro-espaco-natureza}.

A aproxima\c{c}\~{a}o entre o cristianismo e o aristotelismo teve um marco importante com a Filosofia de S\~{a}o Tom\'{a}s de Aquino. De fato, a partir do s\'{e}culo XIII, o aris\-to\-telismo penetrou de forma profunda no pensamento escol\'{a}stico, marcando-o definitivamente. Isso se deveu materialmente \`{a} descoberta de muitas obras de Arist\'{o}teles, e \`{a} tradu\c{c}\~{a}o para o latim de algumas delas, diretamente do grego. Sob a pena de S\~{a}o Tom\'{a}s de Aquino, inicia-se uma tentativa (bem-sucedida) de harmonizar a f\'{e} crist\~{a} e a raz\~{a}o. Para isso, S\~{a}o Tom\'{a}s dedica-se a ``aparar as arestas'' que os ensinamentos aristot\'{e}licos poderiam oferecer \`{a} doutrina da Igreja, uma vez que tecnicamente Arist\'{o}teles era um pag\~{a}o. Essa tend\^{e}ncia foi parte importante da especula\c{c}\~{a}o filos\'{o}fica da \'{e}poca. Em particular, a ideia aristot\'{e}lica de um Cosmos finito no qual a Terra era o centro do Universo era atraente para a Igreja~\cite{Eliade}, talvez por representar um todo finito criado por Deus, o qual poderia bem ser controlado pela Igreja, al\'{e}m, \'{e} claro, de colocar o homem medieval -- que se v\^{e} como feito \`{a} imagem e semelhan\c{c}a de Deus -- no Centro do Universo. Logo, aos poucos, a F\'{\i}sica aristot\'{e}lica foi se transformando em um dogma defendido pela Igreja cat\'{o}lica, n\~{a}o deixando espa\c{c}o para que se questionasse a autoridade do Estagirita, o que seria entendido tamb\'{e}m como um ataque \`{a} institui\c{c}\~{a}o Igreja.

O caminho para a emancipa\c{c}\~{a}o do aristotelismo foi longo. Lynn White, Jr. destaca o fato de que a Ci\^{e}ncia Moderna \'{e} devedora de um r\'{a}pido avan\c{c}o nesse sentido durante os s\'{e}culos XIII e XIV~\cite{Lynn-White}. Segundo ele, a compreens\~{a}o da Filosofia de Arist\'{o}teles levou a cr\'{\i}ticas e a um conhecimento novo de tal modo que n\~{a}o se sustentaria, em seu entendimento, a ideia de que a ci\^{e}ncia moderna seria uma con\-ti\-nua\c{c}\~{a}o pura e simples da ci\^{e}ncia da antiguidade que havia sido interrompida. H\'{a} uma s\'{e}rie de novos ingredientes devidos ao pensamento cr\'{\i}tico desse per\'{\i}odo. Esse ponto de vista \'{e} compartilhado por v\'{a}rios autores, dentre os quais a fil\'{o}sofa alem\~{a} Anneliese Maier, para quem a Hist\'{o}ria da Ci\^{e}ncia entre os s\'{e}culos XIII e XVII \'{e} uma hist\'{o}ria da gradual supera\c{c}\~{a}o do aristotelismo~\cite{Maier}. Essa autora \'{e} de opini\~{a}o que n\~{a}o houve uma \'{u}nica revolu\c{c}\~{a}o nessa supera\c{c}\~{a}o e tampouco ela se deu de forma cont\'{\i}nua e progressiva, conforme sustenta o historiador da ci\^{e}ncia franc\^{e}s Pierre Duhem~\cite{Duhem}. A conclus\~{a}o de Meier \'{e} que houve duas grandes fases em tal processo de mudan\c{c}a: a primeira no s\'{e}culo XIV e a segunda no s\'{e}culo XVII.

Para al\'{e}m dessa discuss\~{a}o, \'{e} importante enfatizar que a supera\c{c}\~{a}o do Cosmos aristot\'{e}lico est\'{a} na base da constru\c{c}\~{a}o de um novo \textit{Cosmos} infinito e homog\^{e}neo, a partir do Renascimento. Al\'{e}m da reestrutura\c{c}\~{a}o da com\-pre\-ens\~{a}o do \textit{movimento}, do \textit{espa\c{c}o}, do \textit{saber} e do \textit{ser},\footnote{\, Lembre-se que todos esses conceitos estiveram envolvidos na Revolu\c{c}\~{a}o das Artes que teve come\c{c}o no s\'{e}culo~XIII e culminou no Renascimento Italiano, como foi explicitado na Se\c{c}\~{a}o~\ref{Arte}.} tal supera\c{c}\~{a}o, como enfatiza Koyr\'{e}, implica e imp\~{o}e a ``reformula\c{c}\~{a}o dos princ\'{\i}pios b\'{a}sicos da raz\~{a}o filos\'{o}fica e cient\'{\i}fica''.

Do ponto de vista de uma hist\'{o}ria internalista da Ci\^{e}ncia, a obra de Cop\'{e}rnico \'{e} considerada um importante divisor de \'{a}guas no de\-sen\-vol\-vimento cient\'{\i}fico. De fato, como afirma Koyr\'{e}~\cite{Koyre},

\begin{quotation}
\noindent \small{O ano de 1543, ano da publica\c{c}\~{a}o do \textit{De Revolutionibus Orbium Coelestium} e da morte do autor, Cop\'{e}rnico, marca uma data importante na hist\'{o}ria do pensamento humano. Estamos tentados a considerar essa data como significando o fim da Idade M\'{e}dia e o come\c{c}o dos tempos modernos, porque, mais que a conquista de Constantinopla pelos turcos ou a descoberta da Am\'{e}rica por Crist\'{o}v\~{a}o Colombo, ela simboliza o fim de um mundo e o co\-me\c{c}o de outro.}
\end{quotation}

\normalsize

Dificilmente poder\'{\i}amos nos referir ao que se seguiu com palavras melhor escolhidas que as de Koyr\'{e}~\cite{Koyre-bis} para sintetizar essa dissolu\c{c}\~{a}o do Cosmos:

\begin{quotation}
\noindent \small{Essa ideia [Cosmos] \'{e} substitu\'{\i}da pela ideia de um Universo aberto, in\-de\-fi\-nido e at\'{e} infinito, unificado e governado pelas mesmas leis universais, um universo no qual todas as coisas pertencem ao mesmo n\'{\i}vel do Ser, contrariamente \`{a} concep\c{c}\~{a}o tradicional que distinguia e opunha os dois mundos do C\'{e}u e da Terra. Doravante, as leis do C\'{e}u e da Terra se fun\-dem. A astronomia e a f\'{\i}sica tornam-se interdependentes,
unificadas e unidas. Isso implica o desaparecimento, da perspectiva cient\'{\i}fica, de todas as considera\c{c}\~{o}es baseadas no valor, na perfei\c{c}\~{a}o, na harmonia, na significa\c{c}\~{a}o e no des\'{\i}gnio. Tais considera\c{c}\~{o}es desaparecem no espa\c{c}o infinito do novo Universo. \'{E} nesse novo Universo, nesse novo mundo, onde uma geometria se faz realidade, que as leis da f\'{\i}sica cl\'{a}ssica encontram valor e explica\c{c}\~{a}o.}
\end{quotation}

\normalsize

Esse \'{e} o Universo que tem origem em Cop\'{e}rnico e, de certa forma, marca o fim da Idade M\'{e}dia. Ele ser\'{a} elaborado por Galileu, que lan\c{c}a as bases da Ci\^{e}ncia Moderna, e, mais tarde, por Newton, que far\'{a} a grande s\'{\i}ntese te\'{o}rica, criando a Teoria da Gravita\c{c}\~{a}o Universal, unificando, ent\~{a}o, formalmente e em car\'{a}ter definitivo, a descri\c{c}\~{a}o f\'{\i}sica dos movimentos na Terra e nos C\'{e}us. Dessa forma, a partir do conceito de \textit{for\c{c}a} e de \textit{espa\c{c}o e tempo absolutos}, Newton estabelece n\~{a}o s\'{o} as causas din\^{a}micas dos movimentos celestes, como Kepler havia tentado alcan\c{c}ar, mas tamb\'{e}m que essa mesma for\c{c}a \'{e} respons\'{a}vel pelos movimentos de queda dos corpos massivos resultantes da atra\c{c}\~{a}o deles pela Terra em sua superf\'{\i}cie.
\newpage

\section{Um pequeno esbo\c{c}o da contribui\c{c}\~{a}o de Galileu \`{a} Astronomia}\label{Galileo}

\textit{Sidereus Nuncius}~\cite{Sidereus} \'{e} um folheto de 24 p\'{a}ginas (Fig.~\ref{fig-caruso-8}) escrito por Galileo Galilei e publicado em Veneza, em mar\c{c}o de 1610.

\begin{figure}[htbp]
\centerline{\includegraphics[width=10.0cm]{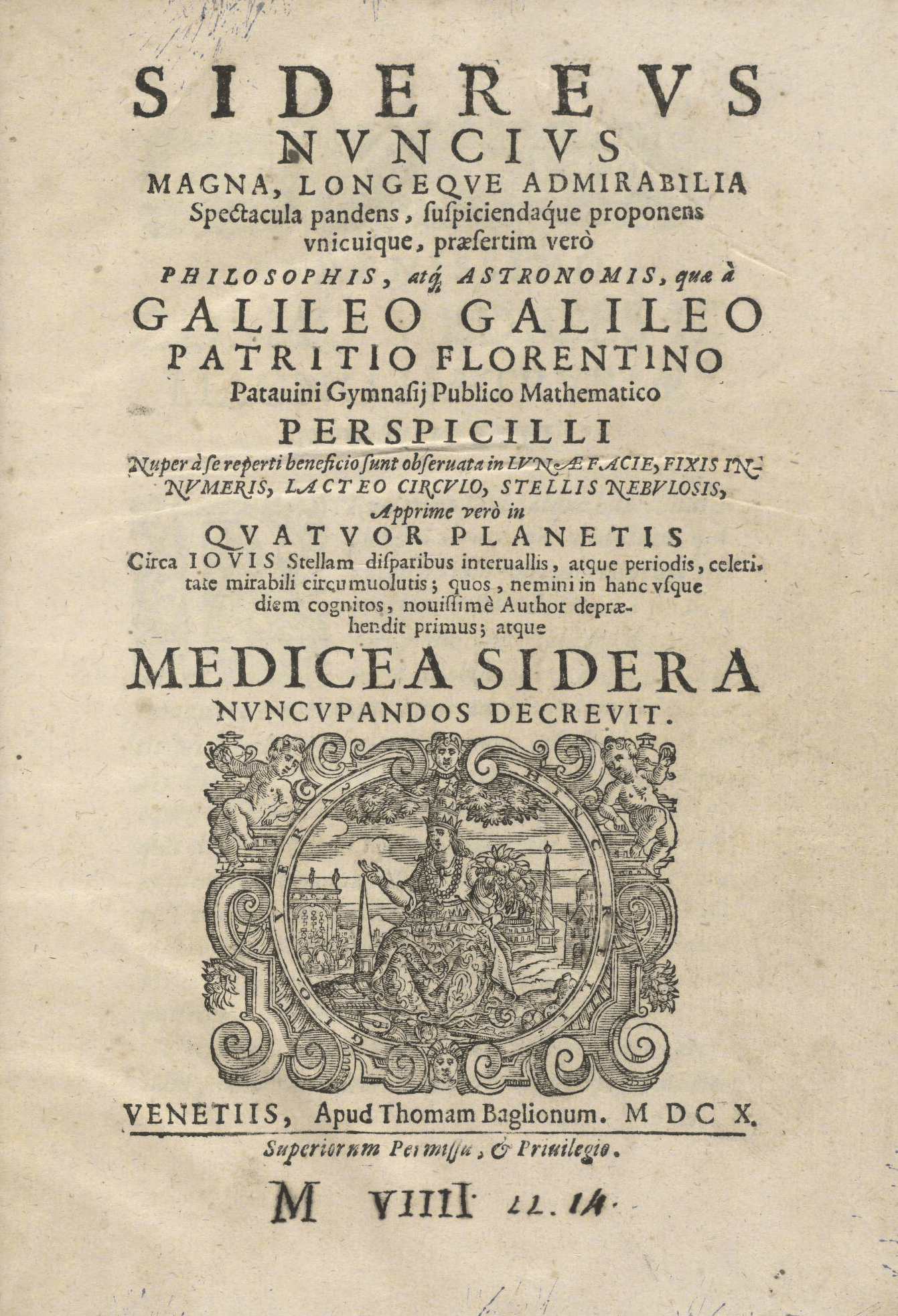}}
\caption{\small{Frontisp\'{\i}cio do livro \textit{Sidereus Nuncius}.}}
\label{fig-caruso-8}
\end{figure}
\newpage

Foi o primeiro tratado cient\'{\i}fico baseado em observa\c{c}\~{o}es astron\^{o}micas rea\-li\-zadas com um telesc\'{o}pio. Cont\'{e}m os resultados das observa\c{c}\~{o}es iniciais da Lua, das estrelas e das luas de J\'{u}piter. Esse folheto foi impresso muito rapidamente, cerca de 2 meses ap\'{o}s suas observa\c{c}\~{o}es.

Nele h\'{a} desenhos feitos pelo pr\'{o}prio Galileu, a partir de suas observa\c{c}\~{o}es, mos\-tran\-do que a superf\'{\i}cie da Lua n\~{a}o \'{e} homog\^{e}nea, pois apresenta crateras, mon\-ta\-nhas e vales (Fig.~\ref{fig-caruso-9}), sugerindo uma semelhan\c{c}a com a Terra. Essa \'{e} uma evi\-d\^{e}n\-cia que con\-tra\-ria a hip\'{o}tese da per\-fei\c{c}\~{a}o do C\'{e}u e de seus astros, como apre\-goa\-va o aris\-to\-telismo.

Al\'{e}m disso, com o aperfei\c{c}oamento da luneta, Galileu constatou que, ao longo do tempo, a face de V\^{e}nus se apresentava iluminada de maneira diferente, indicando que o planeta tem fases, como a Lua. Se V\^{e}nus girasse em torno da Terra, mais pr\'{o}ximo do que o Sol, as suas fases teriam de ser id\^{e}nticas \`{a}s da Lua (Fig.~\ref{fig-caruso-10}), o que n\~{a}o foi verificado. Da\'{\i} o f\'{\i}sico pisano concluiu que o astro realiza um movimento de transla\c{c}\~{a}o ao redor do Sol -- e n\~{a}o em torno da Terra, como a teoria geoc\^{e}ntrica defendia.

\begin{figure}[htbp]
\centerline{\includegraphics[width=7.0cm]{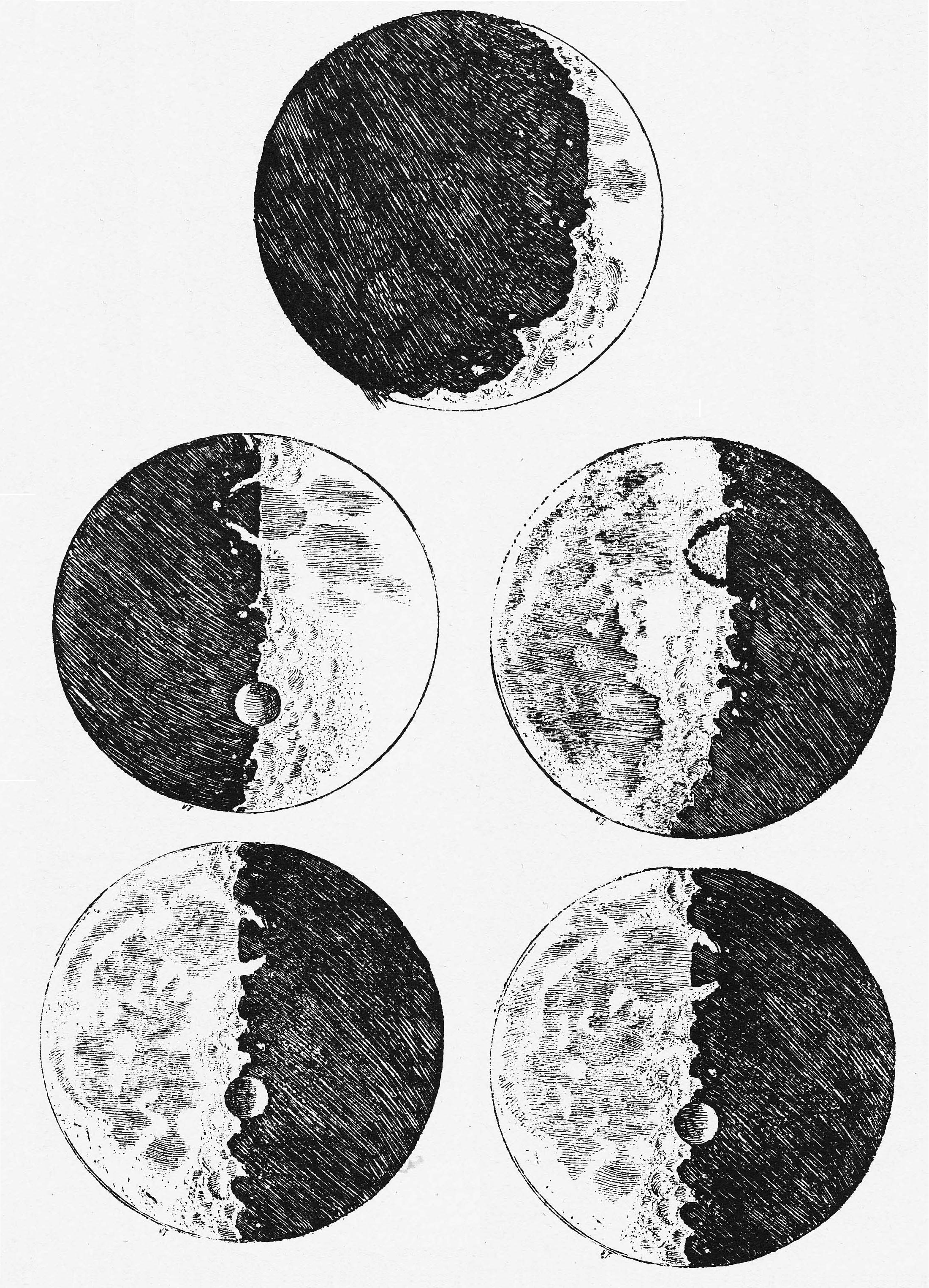}}
\caption{\small{Desenhos da Lua feitos pelo pr\'{o}prio Galileu em seu livro \textit{Sidereus Nuncius}, nos quais destacam-se as crateras e montanhas.}}
\label{fig-caruso-9}
\end{figure}

\newpage
\begin{figure}[htbp]
\centerline{\includegraphics[width=7.0cm]{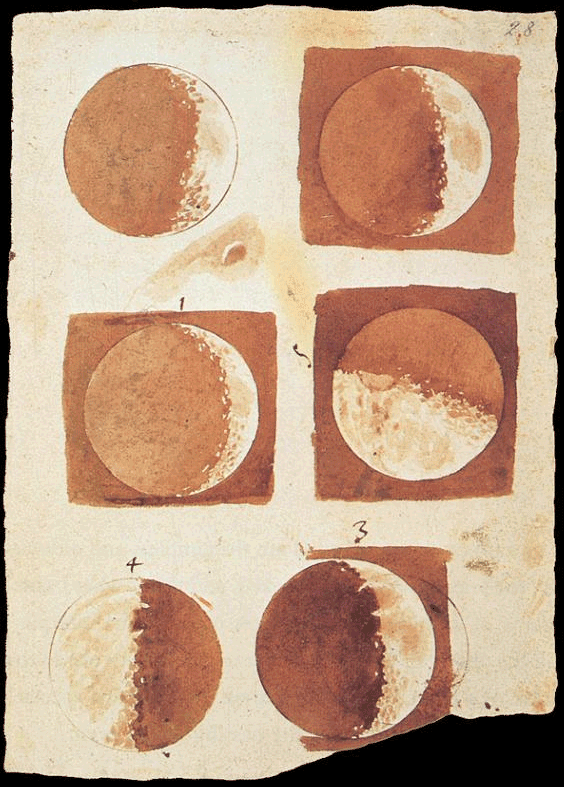}}
\caption{\small{Fases da Lua desenhadas pelo pr\'{o}prio Galileu. Este desenho em s\'{e}pia teria sido feito antes da publica\c{c}\~{a}o de seu \textit{Nuncius Siderius} e foi encontrado junto a uma c\'{o}pia manuscrita do livro em poder de Galileu. O original encontra-se hoje na Biblioteca Nacional de Floren\c{c}a, It\'{a}lia.}}
\label{fig-caruso-10}
\end{figure}

\section{O confronto de ideias e o caminho para a condena\c{c}\~{a}o}

Esta \'{e}, em s\'{\i}ntese, a cronologia dos principais passos de Galileu, colaboradores e de seus opositores durante o longo processo de persegui\c{c}\~{a}o/julgamento/con\-de\-na\c{c}\~{a}o do f\'{\i}sico pisano:

\begin{itemize}
  \item Em 1609, Galileu apontou pela primeira vez uma luneta para o c\'{e}u com o prop\'{o}sito de investig\'{a}-lo cientificamente.
  \item Em 1610 \'{e} publicado o livreto \textit{Sidereus Nuncius}, com as primeiras obser\-va\c{c}\~{o}es de Galileu.
  \item Em 21 de dezembro de 1613, Galileu escreve ao amigo e padre Benedetto Castelli, matem\'{a}tico na Universidade de Pisa, defendendo que as refer\^{e}ncias b\'{\i}blicas aos eventos astron\^{o}micos n\~{a}o deveriam ser tomadas como literais porque os escribas as teriam simplificado para serem compreendidas pelos comuns. Ele tamb\'{e}m afirma que o modelo helioc\^{e}ntrico de Cop\'{e}rnico n\~{a}o era incompat\'{\i}vel com a B\'{\i}blia~\cite{Galileu-cartas}.
  \item Em N\'{a}poles, no ano de 1615, Paolo Antonio Foscarini, um padre e cientista carmelita, publica uma carta aberta com o t\'{\i}tulo \textit{Sopra l'opinione de' Pittagorici, e del Copernico. Della mobilit\`{a} della terra, e stabilit\`{a} del sole, e del nuovo pittagorico sistema del mondo}. Pela primeira vez um cl\'{e}rigo defende abertamente o copernicanismo.
  \item Nesse mesmo ano, o humanista e padre Giovanni Ciampoli, aliado de Ga\-li\-leu, lhe escreve demonstrando sua preocupa\c{c}\~{a}o com a publica\c{c}\~{a}o do livro de Foscarini, posto sob escrut\'{\i}nio pela Inquisi\c{c}\~{a}o.
  \item Ainda no come\c{c}o desse ano, Foscarini enviou uma c\'{o}pia de seu livro sobre a mobilidade da Terra junto com uma carta ao cardeal Roberto Belarmino, obra esta que viria a ser condenada mais tarde pela Inquisi\c{c}\~{a}o Romana, juntamente com o livro de Nicolau Cop\'{e}rnico.
  \item O cardeal Belarmino responde a Foscarini com uma carta datada de 12 de abril de 1615, e envia uma c\'{o}pia a Galileu, que \'{e} citado nominalmente na carta. Nela, o cardeal adverte aos dois ser prudente que ambos se limitem a tratar o heliocentrismo como um fen\^{o}meno meramente hipot\'{e}tico e n\~{a}o fisicamente real. Al\'{e}m disso, acautela-os que interpretar o heliocentrismo como fisicamente real seria ``uma coisa muito perigosa, provavelmente n\~{a}o apenas para irritar todos os fil\'{o}sofos e te\'{o}logos escol\'{a}sticos, mas tamb\'{e}m por prejudicar a F\'{e} Sagrada ao tornar a Sagrada Escritura como falsa''.
  \item Em 1615, Galileu escreve um ensaio tentando acomodar o copernicanismo com as doutrinas da Igreja Cat\'{o}lica e o envia como carta a Madame Cristina de Lorena, Gr\~{a}-duquesa da Toscana. Nela, Galileu argumenta que a teoria copernicana n\~{a}o era apenas uma ferramenta de c\'{a}lculo matem\'{a}tico, mas uma realidade f\'{\i}sica. Essa carta, junto com a enviada a Castelli, em 1613, e alguns outros textos (escritos entre dezembro de 1613 e o final de 1615) foram traduzidos e coligidos em um livro publicado pela Unesp~\cite{Galileu-cartas}.
  \item Em um relat\'{o}rio de 24 de fevereiro de 1616, a Inquisi\c{c}\~{a}o emite um parecer contr\'{a}rio ao heliocentrismo.
  \item Em 5 de maio de 1616, a obra maior de Cop\'{e}rnico \'{e} inclu\'{\i}da no \textit{Index li\-bro\-rum prohibitorum}\footnote{Criado em 1559, pelo Papa Paulo IV, trata-se de um \'{\i}ndice contendo a ``lista negra'' de livros, segundo a Igreja, ou seja, das obras proibidas (e banidas) aos crist\~{a}os por apresentarem conte\'{u}dos teol\'{o}gicos contr\'{a}rios (na avalia\c{c}\~{a}o da Inquisi\c{c}\~{a}o) ao Cristianismo. Esse \textit{Index} s\'{o} foi suprimido, pasme o leitor, em 1966, pelo Conc\'{\i}lio do Vaticano II.} pela Sagrada Congrega\c{c}\~{a}o do Vaticano.
  \item Em outubro de 1623, Galileu escreve \textit{Il Saggiatore}~\cite{Saggiatore}. Embora pretenda tratar de uma quest\~{a}o referente a um cometa, na qual, sabe-se hoje, a posi\c{c}\~{a}o de Galileu era equivocada, essa obra ficou famosa pela defesa galileana de que a Matem\'{a}tica deve ser a linguagem da F\'{\i}sica.
  \item Aos 60 anos, Galileu come\c{c}a a escrever sua grande obra sobre Cosmologia, \textit{Di\'{a}logo sobre os sistemas m\'{a}ximos}, publicados em 1632.
  \item Em 12 de abril de 1633, iniciou-se o julgamento de Galileu~\cite{Westfall} por crime de heresia~\cite{Santillana}.
  \item Com base em uma senten\c{c}a da Inquisi\c{c}\~{a}o, emitida em 22 de junho de 1633, a Igreja Cat\'{o}lica condenou Galileu Galilei por heresia, ap\'{o}s uma batalha de quase 20 anos entre o astr\^{o}nomo italiano e o Vaticano~\cite{Fantoli}. Sua pena foi comutada para pris\~{a}o domiciliar, na qual permanecer\'{a} por todo o resto de sua vida.
  \item Impedido de se dedicar \`{a} Astronomia, Galileu come\c{c}a a se dedicar aos es\-tu\-dos de Mec\^{a}nica e queda dos corpos, formalizados, mais tarde (1638), em sua famosa obra \textit{Discorsi e dimostrazioni matematiche intorno \`{a} due nuove scienze attenenti alla mechanica \& i movimenti locali}~\cite{Duas-ciencias}.
\end{itemize}


\section{\textit{Viri Galilei, quid statis adspicientes in coelum?}}\label{Viri}

O que estava em jogo para Galileu quando ele apontou uma luneta para o C\'{e}u? Era sua inten\c{c}\~{a}o por \`{a} prova o geocentrismo aristot\'{e}lico e justificar o he\-liocentrismo copernicano? Por mais instigante que sejam essas quest\~{o}es, di\-fi\-cil\-mente ter\'{\i}amos como respond\^{e}-las sem recorrer a especula\c{c}\~{o}es. Essa in\-da\-ga\c{c}\~{a}o n\~{a}o \'{e} nova; na verdade, quase t\~{a}o antiga quanto esse ato de Galileu. De fato, em dezembro de 1614, o frade dominicano Tommaso Caccini dirigiu-se em p\'{u}blico a ele e seus seguidores, com uma pergunta similar, escolhida como t\'{\i}tulo dessa Se\c{c}\~{a}o.\footnote{Cuja tradu\c{c}\~{a}o \'{e}: ``Homens da Galileia, por que estais a\'{\i} a olhar para o c\'{e}u?''} Acreditamos que seria mais prudente e eficaz substitu\'{\i}-la por outra diferente, que permita uma res\-pos\-ta mais objetiva, baseada em fatos, testemunhos escritos ou na pr\'{o}pria Obra do f\'{\i}sico pisano.

Por exemplo, podemos nos perguntar o que realmente estava sendo colocado em jogo por Galileu a partir das observa\c{c}\~{o}es colhidas com sua luneta. Ou seja, seu problema com a Igreja resumiu-se \`{a} sua posi\c{c}\~{a}o intransigente com rela\c{c}\~{a}o ao debate sobre o heliocentrismo? A resposta mais frequente a tal pergunta \'{e} \textit{sim}! Al\'{e}m do ataque \`{a} autoridade de Arist\'{o}teles, contido em sua posi\c{c}\~{a}o, ele estaria contribuindo para o que o psicanalista austr\'{\i}aco Sigmund Freud chamou de \textit{a primeira ferida narc\'{\i}sica}~\cite{Freud}, assim chamada por ferir o \textit{ego} humano. O ponto aqui \'{e} que o homem, que se acreditava feito \`{a} imagem e semelhan\c{c}a de Deus e pensava ser um ocupante pri\-vi\-le\-giado do centro do Universo (a Terra), torna-se, de repente, um ser perif\'{e}rico, \`{a} medida que a Terra passa a ser vista t\~{a}o somente como um dos planetas que giram ao redor do Sol. O impacto direto disso sobre o imagin\'{a}rio coletivo de uma sociedade eminentemente religiosa e teoc\^{e}ntrica n\~{a}o \'{e} dif\'{\i}cil de imaginar. Seria essa, ent\~{a}o, toda a causa e ess\^{e}ncia do problema? Ou havia ainda mais coisas envolvidas nessa disputa -- do ponto de vista cognitivo, epistemol\'{o}gico ou mesmo do discurso sobre a Natureza -- capazes de incomodar a Igreja? N\~{a}o teria sido o heliocentrismo um amplo campo de batalha no qual Galileu, ao se posicionar firmemente em sua defesa, recorre a outros pontos de vista (e argumentos) originais e essenciais para a sua concep\c{c}\~{a}o de Mundo (e do que deveria ser o m\'{e}todo ci\-en\-t\'{\i}\-fico), os quais seriam igualmente (ou ainda mais) inaceit\'{a}veis para a Igreja Romana? Tendemos a responder afirmativamente a essa quest\~{a}o como veremos a seguir.

Nossa inten\c{c}\~{a}o a longo prazo \'{e} fazer um estudo hist\'{o}rico-filos\'{o}fico sistem\'{a}tico no sentido de lan\c{c}ar luz sobre essas indaga\c{c}\~{o}es. Como tal estudo est\'{a} apenas co\-me\-\c{c}ando, nos limitamos aqui a ``pensar alto'' e a destacar alguns pontos que nos parecem relevantes, al\'{e}m de tecer alguns coment\'{a}rios sobre nossas conclus\~{o}es preliminares. Em princ\'{\i}pio, ficaremos circunscritos \`{a}s car\-tas es\-critas por Galileu, nas quais ele se defendia das acusa\c{c}\~{o}es de heresia e algumas publica\c{c}\~{o}es entre 1613 e seu julgamento, al\'{e}m do que escreveu em \textit{O Ensa\'{\i}sta}.

Por\'{e}m, antes de prosseguirmos, recordemos um enfoque muito diferente com rela\c{c}\~{a}o ao processo de Galileu~\cite{Documenti-processo}. Segundo o historiador italiano Pietro Redondi, a concep\c{c}\~{a}o atom\'{\i}stica da mat\'{e}ria aceita por Galileu (e n\~{a}o sua defesa do he\-lio\-centrismo) estaria na origem de seu problema com a Inquisi\c{c}\~{a}o~\cite{Redondi}. Sua tese pol\^{e}mica baseia-se na premissa de que, uma vez que os \'{a}tomos s\~{a}o imut\'{a}veis, indestrut\'{\i}veis e indivis\'{\i}veis, fica logicamente impossibilitada a transforma\c{c}\~{a}o do p\~{a}o e do vinho, respectivamente, na carne e no sangue de Cristo, pondo, por conseguinte, em xeque um importante dogma da Igreja: a \textit{eucaristia}.\footnote{\, Supondo plaus\'{\i}vel a tese de Redondi, \'{e} l\'{\i}cito imaginar que a concep\c{c}\~{a}o atual de par\-t\'{\i}culas elementares, na qual a elementaridade n\~{a}o est\'{a} mais relacionada \`{a} indestrutibilidade dos cons\-ti\-tuintes \'{u}ltimos da mat\'{e}ria, pode ter sido levada em conta na revis\~{a}o do processo de heresia de Galileu e ter contribu\'{\i}do, de alguma forma, para a sua absolvi\c{c}\~{a}o, por parte do Vaticano.} At\'{e} onde sabemos, sua tese n\~{a}o foi aceita pela maioria dos historiadores da Ci\^{e}ncia~\cite{Resenha-Westfall,Boido}.

No quadro-resumo da Fig.~\ref{fig-caruso-11} colocamos, lado a lado, alguns t\'{o}picos que mar\-ca\-ram a diferen\c{c}a entre a posi\c{c}\~{a}o eclesi\'{a}stica e a galileana.

Nos albores do s\'{e}culo~XVII, a concep\c{c}\~{a}o que a Igreja tinha com refer\^{e}ncia \`{a} As\-tro\-nomia ainda est\'{a} arraigada ao imagin\'{a}rio e \`{a} representa\c{c}\~{a}o do c\'{e}u dourado. Assim sendo, no pensamento escol\'{a}stico, a Astronomia serve apenas para se ter cons\-ci\^{e}ncia e venerar a obra divina. O C\'{e}u deve ser admirado e apreciado, como mencionado na Se\c{c}\~{a}o~\ref{Arte}. O \'{\i}cone dessa rela\c{c}\~{a}o homem-c\'{e}u poderia ser os \textit{olhos}, com os quais o homem v\^{e} e extasia-se com a Obra divida. Essa determina\c{c}\~{a}o, claramente, \'{e} insepar\'{a}vel da ideia de \textit{verdade absoluta}, que se relaciona ao \textit{eterno}, como veiculada pelo Livro Sagrado (a Sagrada Escritura). J\'{a} para Galileu, o \'{\i}cone dessa mesma rela\c{c}\~{a}o seria o \textit{c\'{e}rebro}, atrav\'{e}s do qual o homem busca uma compreens\~{a}o racional do mundo que o cerca, incluindo o C\'{e}u.

\begin{figure}[htbp]
\centerline{\includegraphics[width=13.0cm]{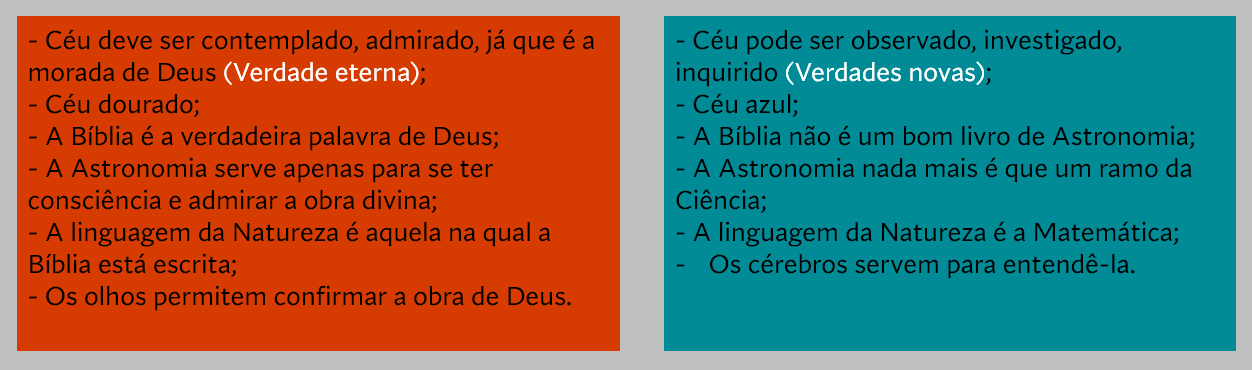}}
\caption{\small{Quadro-resumo dos principais aspectos que caracterizavam e embasavam as vis\~{o}es antag\^{o}nicas da Igreja (quadro laranja) e de Galileu (quadro azul).}}
\label{fig-caruso-11}
\end{figure}

Esta caracter\'{\i}stica nos remete, como foi dito na Se\c{c}\~{a}o~\ref{Arte}, ao fato de que, a partir da revolu\c{c}\~{a}o que tem in\'{\i}cio na Arte com Giotto, a ``ideologia bizantina do eterno'' passa a ser substitu\'{\i}da pela ``ideologia da hist\'{o}ria'', como apropriadamente afirmou Giulio Argan. Ainda usando suas pa\-la\-vras, essa nova atitude contribui para formar ``o pensamento de que a cons\-ci\^{e}ncia da hist\'{o}ria seja a base de todo interesse cognitivo e \'{e}tico''. E a Hist\'{o}ria, constru\'{\i}da pelo homem, inevitavelmente, nos leva a \textit{novas verdades}. As verdades podem, por conseguinte, ser \textit{provis\'{o}rias}. Logo, ao aceitar que a investiga\c{c}\~{a}o cient\'{\i}fica \'{e} capaz de conduzir a novas verdades a partir de um processo cognitivo humano envolvendo, como defendia Bacon (Se\c{c}\~{a}o~\ref{Optica}), a observa\c{c}\~{a}o, a experimenta\c{c}\~{a}o, a elabora\c{c}\~{a}o de hip\'{o}teses e a ne\-ces\-sidade de verifica\c{c}\~{a}o independente, vemos, na atitude inicial de Galileu,\footnote{Sua originalidade foi incorporar ainda a Matem\'{a}tica como linguagem da F\'{\i}sica e da Astronomia.} \textit{mutatis mutandis}, uma postura muito semelhante \`{a}quela de Giotto no s\'{e}culo~XIII. Ele admite, enfim, que a \textit{Verdade} deva ser buscada no \textit{Livro da Natureza} e n\~{a}o t\~{a}o somente na B\'{\i}blia, quando o assunto \'{e} o mundo real que nos cerca. Cria-se uma tens\~{a}o entre os dois Livros. Estamos observando um per\'{\i}odo hist\'{o}rico no qual se nota, usando a express\~{a}o de Cassirer, o aparecimento de um novo \textit{estado de tens\~{a}o} do pensamento~\cite{Cassirer-cita}, caracter\'{\i}stico do in\'{\i}cio do Renascimento. Dito de outra forma:

\begin{quotation}
\noindent \small{Assim como as artes pl\'{a}sticas buscam f\'{o}rmulas pl\'{a}sticas de concilia\c{c}\~{a}o, tamb\'{e}m a filosofia busca f\'{o}rmulas conceituais de concilia\c{c}\~{a}o ``entre a con\-fi\-an\c{c}a medieval em Deus e a confian\c{c}a do homem do Renascimento em si mesmo''~\cite{Cassirer-cita}.}
\end{quotation}

\normalsize

Mas esta concilia\c{c}\~{a}o vai ter seu pre\c{c}o. De outra parte, quanto ao processo de seculariza\c{c}\~{a}o da Ci\^{e}ncia, ainda segundo Cassirer, ele ``se consuma no momento em que, mais tarde, a revela\c{c}\~{a}o do `livro da natureza' \'{e} contraposta \`{a} revela\c{c}\~{a}o da B\'{\i}blia'', como efetivamente far\'{a} Galileu Galilei.

Chegamos aqui a um ponto crucial, para o qual Galileu deu uma contribui\c{c}\~{a}o original e fundamental, de grande impacto no desenvolvimento do conhecimento cient\'{\i}fico, mas que lhe custou a sua liberdade. Em primeiro lugar, \'{e} preciso en\-ten\-der que n\~{a}o pode haver qualquer oposi\c{c}\~{a}o de princ\'{\i}pio entre o \textit{Livro das Sa\-gra\-das Escrituras} e o \textit{Livro da Natureza}. \'{E} essa equival\^{e}ncia, no fundo, que garante a velha cren\c{c}a de que ambos s\~{a}o livros capazes de nos levar a Deus. \'{E} ela que reflete a unidade do criador divino na natureza. Aceito isso, a dificuldade surge quando, ainda que seja aparentemente, nos confrontamos com alguma contradi\c{c}\~{a}o entre ambos. Sobre este ponto espec\'{\i}fico, na carta a Dom Benedetto Castelli~\cite{Galileu-cartas}, Galileu afirma ``que a Sagrada Escritura n\~{a}o pode nunca mentir ou errar, mas serem os seus decretos de absoluta e inviol\'{a}vel verdade''. E prossegue dizendo que

\begin{quotation}
\noindent \small{S\'{o} teria acrescentado que, se bem a Escritura n\~{a}o pode errar, n\~{a}o me\-nos poderia \`{a}s vezes errar alguns dos seus int\'{e}rpretes e expositores, de v\'{a}\-rios modos. Entre estes, um seria muit\'{\i}ssimo grave e frequente; quando qui\-ses\-se deter-se sempre no puro significado das palavras; porque, assim, apa\-re\-ce\-riam a\'{\i} n\~{a}o apenas diversas contradi\c{c}\~{o}es, mas graves heresias e mesmo blasf\^{e}mias.}
\end{quotation}
\normalsize

Esse tipo de tentativa de contornar um problema incontorn\'{a}vel, leva Galileu, sucessivamente, a apresentar, em sua defesa, argumentos semelhantes a esses em suas cartas, \`{a}s vezes parecendo-nos verdadeiros sofismas. Claro que a Hist\'{o}ria nos mostra que foram todos infrut\'{\i}feros. Galileu havia introduzido uma assimetria irremedi\'{a}vel entre o \textit{Livro das Escrituras} e o \textit{Livro da Natureza}. Nessa v\~{a} ten\-ta\-tiva de defesa, levantou claramente a limita\c{c}\~{a}o da \textit{palavra}. Em \'{u}ltima an\'{a}lise, ele confere~\cite{Cassirer-cita-2}

\begin{quotation}
\noindent \small{\`{a} revela\c{c}\~{a}o contida na obra [no \textit{Livro da Natureza}] a primazia em rela\c{c}\~{a}o \`{a} palavra: pois a palavra \'{e} algo que existe e perdura como algo que, direta e presentemente, se nos oferece \`{a} consulta.}
\end{quotation}
\normalsize

Portanto, ele est\'{a} admitindo que a linguagem da B\'{\i}blia n\~{a}o \'{e} apropriada \`{a} Ci\^{e}n\-cia, al\'{e}m de explicitar, em algumas cartas, que o prop\'{o}sito da B\'{\i}blia n\~{a}o \'{e} abso\-lu\-tamente dar uma descri\c{c}\~{a}o detalhada do Mundo, at\'{e} mesmo porque n\~{a}o \'{e} sua verdadeira inten\c{c}\~{a}o.

Cop\'{e}rnico tira o homem do centro do Mundo, mas Galileu o recoloca, de cer\-ta forma, no centro; mais especificamente, no centro do \textit{Livro da Natureza}. Nes\-se caminho, o f\'{\i}sico pisano torna-se um cr\'{\i}tico da palavra, da autoridade, e do discurso b\'{\i}blico em torno \`{a} Natureza. A linguagem calcada na \textit{palavra}, mais suscet\'{\i}vel a in\-ter\-pre\-ta\c{c}\~{o}es, deve ser substitu\'{\i}da por outra, mais formal e me\-nos amb\'{\i}gua: a \textit{Matem\'{a}tica}. Essa seria a linguagem capaz de levar a cons\-tru\-\c{c}\~{o}es sint\'{a}ticas mais ri\-go\-rosas e mais livres da opini\~{a}o, da \textit{doxa}, tornando-se, idealmente, imune \`{a}s disputas teol\'{o}gicas.

De fato, em 1623, dez anos antes da sua condena\c{c}\~{a}o, Galileu formaliza o que se pode chamar de seu elogio \`{a} Ma\-te\-m\'{a}tica como linguagem do universo f\'{\i}sico, no seu novo livro \textit{Il Saggiatore}. Sua ideia sobre o papel da Matem\'{a}tica pode ser sintetizado pela seguinte cita\c{c}\~{a}o~\cite{Saggiatore}:

\begin{quotation}
\noindent \small{A filosofia est\'{a} escrita neste grande livro continuamente aberto para os nossos olhos (quero dizer, o universo), mas que n\~{a}o pode ser compreendido sem antes aprendermos a compreender a l\'{\i}ngua, conhecer os caracteres nos quais est\'{a} escrito. Ele \'{e} escrito em linguagem matem\'{a}tica, e os ca\-rac\-teres s\~{a}o tri\-\^{a}n\-gulos, c\'{\i}rculos e outras figuras geom\'{e}tricas, sem as quais \'{e} hu\-ma\-na\-mente imposs\'{\i}vel entender uma palavra: sem estes se estar\'{a} vagando num labirinto escuro.}
\end{quotation}

\normalsize

Com isso, Galileu atribui um significado um pouco diferente \`{a} \textit{observa\c{c}\~{a}o}. Para ele \'{e} preciso, segundo Cassirer, que o esp\'{\i}rito humano seja capaz de relacionar o conte\'{u}do, o fruto de sua percep\c{c}\~{a}o, \`{a}quelas formas b\'{a}sicas do conhecimento, cuja imagem primordial ele deve carregar consigo. ``Somente gra\c{c}as a essa rela\c{c}\~{a}o e a essa interpreta\c{c}\~{a}o \'{e} que o livro da natureza se nos torna leg\'{\i}vel e compreens\'{\i}vel''~\cite{Cassirer-cita-3}. \'{E} preciso observar, fazer hip\'{o}teses, traduzir o fen\^{o}meno matematicamente e confrontar os resultados esperados com a experi\^{e}ncia.

Percebemos at\'{e} aqui que est\~{a}o impl\'{\i}citas, tamb\'{e}m no trabalho de Galileu, tr\^{e}s caracter\'{\i}sticas iguais ou an\'{a}logas \`{a}quelas que destacamos na revolu\c{c}\~{a}o da arte pict\'{o}rica do s\'{e}culo~XIII que se originou com Giotto. S\~{a}o elas: a valoriza\c{c}\~{a}o do indiv\'{\i}duo e da hist\'{o}ria; a tentativa de seculariza\c{c}\~{a}o n\~{a}o da Arte mas da F\'{\i}sica e da Astronomia; e a busca por uma re\-pre\-senta\c{c}\~{a}o mais real da Natureza.

Outra caracter\'{\i}stica daquele per\'{\i}odo apontada na Se\c{c}\~{a}o~\ref{Arte} foi a Geometriza\c{c}\~{a}o da Pintura. No que concerne \`{a} Astronomia, j\'{a} mencionamos as contribui\c{c}\~{o}es de Cop\'{e}rnico e Kepler (Se\c{c}\~{a}o~\ref{Cosmos}). Mas h\'{a} que se mencionar tamb\'{e}m Galileu. De fato, ao dedicar-se \`{a} quest\~{a}o do \textit{movimento} depois de sua pris\~{a}o domicilar, Galileu recorre \`{a} Geometria para estudar e descobrir as leis de movimento na Terra. Didaticamente, esse tema foi abordado em~\cite{Galileu-sala-de-aula}.

Deixamos por \'{u}ltimo a tentativa de representar o movimento, tamb\'{e}m presente no trabalho de Galileu, pois ela \'{e} tratada tardiamente em sua obra. Na Astronomia, j\'{a} mencionamos a tentativa de Kepler de explicar fisicamente o movimento dos planetas. Na verdade, essa busca faz parte de uma tend\^{e}ncia que permeia todo o Renascimento, ou seja, o desejo de entender a Natureza a partir de seus pr\'{o}prios princ\'{\i}pios. E ao se projetar esse desejo \`{a} descri\c{c}\~{a}o do movimento dos c\'{e}us, esbarramo-nos num s\'{e}rio obst\'{a}culo herdado do pensamento medieval, a saber, a ideia de que h\'{a} um destino pr\'{e}-estabelecido para o Mundo e que nosso destino est\'{a}, de alguma forma, escrito nos c\'{e}us, cabendo \`{a} Astrologia ler e decifrar os signos desses destinos. Esse ponto \'{e} importante~\cite{livro-espaco-natureza} pois a doutrina crist\~{a}, que permeou e dominou a Idade M\'{e}dia no Ocidente, baseia-se no pressuposto da exist\^{e}ncia de uma provid\^{e}ncia geral regendo o Mundo e os destinos do Homem. \'{E}, por consequ\^{e}ncia, um per\'{\i}odo no qual a Astrologia desempenha uma fun\c{c}\~{a}o central na sociedade medieval~\cite{Thorndike}. Foi Kepler quem abriu a porta para que a credibilidade da Astronomia superasse a da Astrologia. Desse momento em diante, abala-se a ideia de uma ordem c\'{o}smica (j\'{a} comprometida pela revolu\c{c}\~{a}o copernicana, como foi visto), dominada por um determinismo divino, o espa\c{c}o geom\'{e}trico come\c{c}a a tomar o lugar do espa\c{c}o hierarquizado, m\'{\i}tico e m\'{a}gico do imagin\'{a}rio medieval, e o homem torna-se mais livre para exercer sua raz\~{a}o e desenvolver uma vis\~{a}o do Mundo e, em particular, do C\'{e}u, menos divina, mais ``humanizada''.

Assim, abre-se caminho para que o estudo dos movimentos dos corpos celestes se livrem de qualquer conota\c{c}\~{a}o m\'{\i}tica e possa se reduzir \`{a} procura de uma ex\-pli\-ca\c{c}\~{a}o em termos de \textit{for\c{c}as} inatas que agem na Natureza. Foi nesse sentido que Kepler buscou uma unifica\c{c}\~{a}o da Astronomia com a F\'{\i}sica, tentando estabelecer a causa din\^{a}mica das \'{o}rbitas dos planetas em torno do Sol, sem sucesso, no entanto.

Como \'{e} bem sabido, Galileu dedicou-se ao estudo matem\'{a}tico dos movimentos locais e publicou seus resultados em seu famoso livro \textit{Duas novas ci\^{e}ncias}~\cite{Duas-ciencias}. A maior dificuldade posta para a F\'{\i}sica, entre meados do s\'{e}culo XVI e in\'{\i}cio do s\'{e}culo XVII, foi compreender o \textit{movimento e suas causas}. Com efeito, dados o espa\c{c}o euclidiano tridimensional e o tempo linear, o cen\'{a}rio do Mundo onde se desenvolveram a F\'{\i}sica e a Hist\'{o}ria como ci\^{e}ncias poss\'{\i}veis de h\'{a} tempos estava posto, mas a compreens\~{a}o do movimento exigiria, um novo esquema causal, uma revolu\c{c}\~{a}o completa na Ci\^{e}ncia e na Cultura, associada ao nome de Isaac Newton, mas devedora tamb\'{e}m a Cop\'{e}rnico, Kepler e Galileu~\cite{livro-espaco-natureza,Jenner}.

Vimos, ao longo do texto, que, para que Galileu pudesse apontar sua luneta para o c\'{e}u, foi necess\'{a}rio, antes, que algu\'{e}m pintasse o c\'{e}u de azul. Nossa contribui\c{c}\~{a}o original aqui refere-se a termos mostrado que quem passou a pintar o c\'{e}u de azul numa cultura que o desejava dourado, Giotto, percorreu um trajeto conceitual semelhante ao trilhado por Galileu s\'{e}culos mais tarde.

Por outro lado, argumentamos que seria uma super simplifica\c{c}\~{a}o atribuir o in\-for\-t\'{u}nio de Galileu com a Igreja pura e simplesmente \`{a} sua defesa do he\-lio\-cen\-tris\-mo. \'{E} claro que ele contraria diretamente algumas passagens da B\'{\i}blia, como aquela na qual Deus teria parado o movimento do Sol a pedido de Josu\'{e}. Mas, esse tipo de contradi\c{c}\~{a}o pode ser visto como de car\'{a}ter mais pontual, tendo Galileu oferecido em sua carta a Dom Benedetto Castelli, por exemplo, alternativa para contornar o problema. O heliocentrismo foi, na realidade, a ponta de um \textit{iceberg}. As consequ\^{e}ncias de ter apontado uma luneta para o c\'{e}u foram muito al\'{e}m desse detalhe. Galileu, a partir de suas observa\c{c}\~{o}es, mas tamb\'{e}m da concep\c{c}\~{a}o pr\'{o}pria que tinha do que \'{e} observar a Natureza e do que deveria ser a Ci\^{e}ncia, introduziu uma assimetria entre os dois Livros: o da Escritura e o da Natureza. Com ela, claramente a Igreja se viu amea\c{c}ada, a partir do momento em que Galileu critica o uso das palavras e do discurso baseado na linguagem comum como meio de descrever a Natureza e seus comportamentos. Al\'{e}m disso, h\'{a} a possibilidade aterradora de que as verdades n\~{a}o sejam mais eternas e, em vista disso, mais cedo ou mais tarde, essa vulnerabilidade do conceito de \textit{Verdade} abalaria (ou pelo menos amea\c{c}aria) os dogmas da Igreja. A defesa da Matem\'{a}tica como linguagem da As\-tro\-nomia e da F\'{\i}sica~\cite{Powers} minaria a autoridade eclesi\'{a}stica, como efetivamente ocorreu, ainda que de forma lenta, mas inexor\'{a}vel.

Einstein referiu-se a quanto Galileu representou uma amea\c{c}a aos dogmas caros \`{a} Igreja com essas palavras~\cite{Einstein}:

\begin{quotation}
\noindent \small{O \textit{leitmotif} que reconhe\c{c}o na obra de Galileu \'{e} a luta apaixonada contra qualquer tipo de dogma baseado na autoridade. Apenas a experi\^{e}ncia e a reflex\~{a}o cuidadosa s\~{a}o aceitos por ele como crit\'{e}rios de verdade. Hoje em dia, nos \'{e} dif\'{\i}cil entender qu\~{a}o sinistra e revolucion\'{a}ria tal atitude pareceu na \'{e}poca de Galileu, quando apenas duvidar da verdade de opini\~{o}es que n\~{a}o tinham base, exceto na autoridade, era considerado um crime capital e punido em conformidade.}
\end{quotation}

\normalsize

Em suma, certamente o frade Tommaso Caccini n\~{a}o poderia ter a menor ideia do amplo impacto transformador que estaria contido na resposta galileana \`{a} sua indaga\c{c}\~{a}o: \textit{Viri Galilei, quid statis adspicientes in coelum?}

\section*{Agradecimentos}

Gostar\'{\i}amos de agradecer a Alessandra Balbi, Felipe Silveira e Roberto Moreira Xavier de Ara\'{u}jo pela ajuda na revis\~{a}o final do texto. A Moreira deixamos tamb\'{e}m nosso reconhecimento por algumas sugest\~{o}es feitas que permitiram uma maior clareza do texto.



\end{document}